\documentclass[aps,floats]{revtex4}
\usepackage{amsmath,amssymb}
\usepackage{graphicx,epsfig}
\usepackage[greek,english]{babel}
\usepackage{bbold}
\usepackage{braket}

\begin{document}
\bibliographystyle {plain}

\pdfoutput=1
\def\oppropto{\mathop{\propto}} 
\def\opsimeq{\mathop{\simeq}}
\def\opoverderline{\mathop{\overline}}
\def\operarrow{\mathop{\longrightarrow}}
\def\opsim{\mathop{\sim}}

\def\fig#1#2{\includegraphics[height=#1]{#2}}
\def\figx#1#2{\includegraphics[width=#1]{#2}}


\title{ Construction of Many-Body-Localized Models  \\
  where all the eigenstates are Matrix-Product-States}


\author{ C\'ecile Monthus }
 \affiliation{Institut de Physique Th\'eorique, 
Universit\'e Paris Saclay, CNRS, CEA,
91191 Gif-sur-Yvette, France}

\begin{abstract}
The inverse problem of  'eigenstates-to-Hamiltonian' is considered for an open chain of $N$ quantum spins in the context of Many-Body-Localization. We first construct the simplest basis of the Hilbert space made of $2^N$ orthonormal Matrix-Product-States (MPS), that will thus automatically satisfy the entanglement area-law. We then analyze the corresponding $N$ Local Integrals of Motions (LIOMs) that can be considered as the local building blocks of these $2^N$ MPS, in order to construct the parent Hamiltonians that have these $2^N$ MPS as eigenstates. Finally we study the Matrix-Product-Operator form of the Diagonal Ensemble Density Matrix that allows to compute long-time-averaged observables of the unitary dynamics. Explicit results are given for the memory of local observables and for the entanglement properties in operator-space, via the generalized notion of Schmidt decomposition for density matrices describing mixed states.

\end{abstract}

\maketitle

\section{ Introduction }

Many-Body-Localization for quantum interacting disordered systems
is a fascinating phase of matter with very unusual properties 
with respect to the standard thermalization scenario of statistical physics (see the reviews 
\cite{nandkishore15,altman15mblreview,parameswaran17,review_mblergo,review_prelovsek,review_rare,alet18,abanin18}
and references therein). In particular, excited eigenstates display an entanglement area-law \cite{bauer,kjall,alet,lim1,lim2}
instead of the entanglement volume-law of thermalized eigenstates.
As a consequence, they can be efficiently approximated by Matrix Product States or DMRG-X algorithms
 \cite{pekker1,pekker2,friesdorf,tensor,sondhi,dmrgx,dmrgxfloquet}
that generalize to excited states the Density-Matrix-RG algorithm concerning ground-states \cite{white1,white2,ulrich2005}.
Another proposal is to construct them via the RSRG-X procedure \cite{pekker14rsrgx,you16,huang14,pouranvari15,agarwal15,
vasseur16particlehole,slage16,friedman17,vasseur15hot,kang17,monthus16emergent,monthus17mblcayley,rsrgxmajorana}
that generalizes to excited states the Strong Disorder Real-Space RG approach  \cite{strong_review}
introduced initially by Ma-Dasgupta-Hu \cite{ma_dasgupta}
 and Daniel Fisher \cite{fisher} to construct the ground states of random quantum spin chains.
Another surprising property is that the Many-Body-Localized phase
can be characterized
 by an extensive number of Local Integrals of Motions (LIOMs)
\cite{emergent_swingle,emergent_serbyn,emergent_huse,
emergent_ent,imbrie,serbyn_quench,emergent_vidal,emergent_ros,emergent_rademaker,
serbyn_powerlawent,c_emergent,ros_remanent,wortis,imbrie17,rademaker17,perlioms,maj_pol,counting_lioms}.
The emergence of these LIOMs can be for instance understood within the RSRG-t procedure 
\cite{vosk13,vosk14,bukov15,monthus17rsrgt,huang17}
that generalizes to the unitary dynamics the Strong Disorder Real-Space RG approach already mentioned above.
These LIOMs can be interpreted as the building blocks of the whole set of eigenstates.

The main activity in the field of Many-Body-Localization has been focused on the 'direct problem',
where one analyzes the properties of a given disordered interacting Hamiltonian $H$
 in order to determine if a Many-Body-Localized phase exists in a certain region of parameters,
usually numerically or via approximate analytical methods,
although some mathematically rigorous results also exist \cite{imbrie,mbllandscape}.
In the present paper, we will instead consider the 'inverse problem' : 
we will first build an orthonormal basis of the Hilbert space made of Matrix-Product-States,
that will thus automatically satisfy the entanglement area-law;
we will then construct the parent Hamiltonians that have these Matrix-Product-States as eigenstates.
This 'inverse problem' perspective is well-known in the field of Tensor Networks
(see the reviews \cite{wolf,ver,cirac,vidal_intro,ulrich2011,phd-evenbly,mera-review,hauru,orus14a,orus14b,orus19} and references therein) in particular to construct local parent Hamiltonians that have a given Matrix-Product-State as groundstate \cite{wolf}, and has also produced the 'eigenstate-to-Hamiltonian' method \cite{ranard,clark}
with recent applications concerning Many-Body-Localized excited states \cite{dupont1,dupont2}
and the engineering of topological models with desired properties \cite{topo}.


The paper focuses on an open chain of $N$ quantum spins and is organized as follows.
In section \ref{sec_MPSbasis}, we introduce the simplest Matrix-Product-States-basis for the Hilbert space of size $2^N$,
 where the $2^N$ orthonormal MPS  have the same entanglement entropy across each bond and have the same multifractal properties.
In section \ref{sec_MPObasis}, we analyze the corresponding Matrix-Product-Operator-basis for the Operator-space of size $4^N$ and we describe the $N$ pseudo-spins that are the building blocks of the $2^N$ MPS.
In section \ref{sec_Hamilton}, we construct the parent Hamiltonians that have these $2^N$ MPS as eigenstates.
In section \ref{sec_dynamics}, we study the properties of the Diagonal Ensemble Density Matrix that allows to compute 
long-time-averaged observables of the unitary dynamics, with explicit results for the memory of local observables and for the entanglement properties in Operator-space.
Our conclusions are summarized in section \ref{sec_conclusion}.
The Appendix \ref{app_majorana} contains the interpretation of the Local Integrals of Motion (LIOMs) in terms
of Majorana fermions.


\section{ An Hilbert space basis made of orthonormal Matrix-Product-States }

\label{sec_MPSbasis}

The notion of entanglement between the different regions of many-body quantum systems
has completely changed the perspective on many condensed-matter problems
(see the reviews \cite{amico08,horo,calabrese09,qi,laflorencie16,chiara} and references therein).
 In particular for one-dimensional quantum spin chains,
the Matrix-Products-States (MPS) are well adapted to describe non-critical states
displaying area-law entanglement \cite{wolf,ver,cirac,vidal_intro,ulrich2011,phd-evenbly,mera-review,hauru,orus14a,orus14b,orus19}.
While this area-law is usually only valid for ground-states \cite{hastings},
we have recalled in the Introduction that the area-law also applies to excited states in Many-Body-Localized phases
\cite{pekker1,pekker2,friesdorf,tensor,sondhi,dmrgx,dmrgxfloquet}.
In this section, our goal is thus to construct the simplest MPS basis for the Hilbert space and to analyze its basic properties.

\subsection{ Simplest Matrix-Product-States-basis for an open chain of $N$ quantum spins }

For an open chain of $N$ quantum spins $j=1,2..,N$,
we write the following $2^N$ Matrix-Product-States of bond dimension $D=2$
\begin{eqnarray}
\ket{\psi_{\epsilon_1,...,\epsilon_{N-1},P  } } && =  \sum_{\alpha_1=\pm} ...  \sum_{\alpha_{N-1}=\pm} 
\left[ \prod_{k=1}^{N-1}\lambda^{[\epsilon_k] \alpha_k}_k  \right]
\ket{\sigma^z_1=  \alpha_1} 
\left[ \prod_{k=2}^{N-1} \ket{\sigma^z_k= \alpha_{k-1} \alpha_k} \right]
\ket{\sigma^z_N=  \alpha_{N-1} P} 
\label{mpsfinal}
\end{eqnarray}
that are labelled by the eigenvalue $P=\pm 1$ 
\begin{eqnarray}
P^z  \ket{\psi_{\epsilon_1,...,\epsilon_{N-1} ,P } }= P \ket{\psi_{\epsilon_1,...,\epsilon_{N-1},P  } }
\label{parityeigen}
\end{eqnarray}
of the Parity operator
\begin{eqnarray}
P^z \equiv \prod_{j=1}^N \sigma_j^z
\label{paritysigmaz}
\end{eqnarray}
and by $(N-1)$ binary variables $\epsilon_1=\pm,...,\epsilon_{N-1}=\pm $
associated to the $(N-1)$ bonds. The variable $\epsilon_k$
determines the two possible Schmidt values $\lambda_k^{[\epsilon_k] \alpha=\pm} $ across the bond $(k,k+1)$ that appear in the MPS of Eq. \ref{mpsfinal} (Eq. \ref{mpsfinal} is written in the Vidal canonical form \cite{vidalcanonical}
where all the Schmidt values appear explicitly, as explained in more details below around Eq. \ref{mpsschmidt})
\begin{eqnarray}
\lambda_k^{[\epsilon] \alpha} && \equiv 
\cos \left( \frac{\theta_k}{2} \right)
\left[ \delta_{\epsilon,+} \delta_{\alpha,+} - \delta_{\epsilon,-}\delta_{\alpha,-} \right]
+ \sin \left( \frac{\theta_k}{2} \right)
\left[ \delta_{\epsilon,-} \delta_{\alpha,+} + \delta_{\epsilon,+}\delta_{\alpha,-} \right]
 \nonumber \\ &&
 = \cos \left( \frac{\theta_k}{2} \right) \epsilon \delta_{\epsilon,\alpha}
+ \sin \left( \frac{\theta_k}{2} \right)  \delta_{\epsilon,-\alpha}
\label{lambdacomplexksum}
\end{eqnarray}

Using the property
\begin{eqnarray}
\sum_{\alpha=\pm}\lambda^{[\epsilon] \alpha}_k 
\lambda^{[\epsilon'] \alpha}_k     =\delta_{\epsilon,\epsilon'}
\label{ortho}
\end{eqnarray}
one can check the orthonormalization of the $2^N$ MPS of Eq. \ref{mpsfinal}
\begin{eqnarray}
\braket{ \psi_{\epsilon_1',...,\epsilon_{N-1}',P'} 
 \vert \psi_{\epsilon_1,...,\epsilon_{N-1},P } } && =
\sum_{ \substack{\alpha_1 =\pm \\ \alpha_1'=\pm} }
...
\sum_{ \substack{\alpha_{N-1} =\pm \\ \alpha_{N-1}'=\pm} }
\left[ \prod_{k=1}^{N-1}\lambda^{[\epsilon_k] \alpha_k}_k 
\lambda^{[\epsilon_k'] \alpha_k'}_k    \right]
\delta_{\alpha_1',\alpha_1} 
\left[ \prod_{k=2}^{N-1} \delta_{\alpha_{k-1}' \alpha_k', \alpha_{k-1} \alpha_k} \right]
\delta_{ \alpha_{N-1}' P',  \alpha_{N-1} P }
\nonumber \\
&& = \delta_{P',P}
\prod_{k=1}^{N-1} \left[ \sum_{\alpha_k=\pm}\lambda^{[\epsilon_k] \alpha_k}_k 
\lambda^{[\epsilon_k'] \alpha_k}_k     \right]
= \delta_{P',P}
\prod_{k=1}^{N-1} \delta_{\epsilon_k',\epsilon_k}
\label{orthonorm}
\end{eqnarray}

Since the $(N-1)$ angles $\theta_k$ for $k=1,..,N-1$ appearing in Eq. \ref{lambdacomplexksum}
are the only free parameters of this simple MPS basis,
it is important to explain now their physical meaning
in terms of the entanglement properties across bonds and in terms of the multifractality
of individual MPS.


\subsection{ Entanglement properties of individual MPS across the bond $(n,n+1)$}

The MPS of Eq. \ref{mpsfinal} are written in the Vidal canonical form \cite{vidalcanonical},
where the entanglement properties for any bi-partitioning of the chain into two parts $[1,..,n]$ and $[n+1,..,N]$
are directly accessible.
The Schmidt decomposition of the ket of Eq. \ref{mpsfinal}
with respect to the bond $[n,n+1]$ 
reads
\begin{eqnarray}
\ket{\psi_{\epsilon_1,...,\epsilon_{N-1},P  } } && = 
 \sum_{\alpha_n=\pm}
\ket{\Phi_{\epsilon_1,...,\epsilon_{n-1}  }^{[1,..,n] \alpha_n} }
\lambda^{[\epsilon_n] \alpha_n}_n 
\ket{\Phi_{\epsilon_{n+1} ..,\epsilon_{N-1},P }^{[n+1,..,N]\alpha_n} }
\label{mpsschmidt}
\end{eqnarray}
where
\begin{eqnarray}
\ket{\Phi_{\epsilon_1,...,\epsilon_{n-1}  }^{[1,..,n] \alpha_n} }  && = 
 \sum_{\alpha_1=\pm} ...  \sum_{\alpha_{n-1}=\pm} 
\left[ \prod_{k=1}^{n-1}\lambda^{[\epsilon_k] \alpha_k}_k  \right]
\ket{\sigma^z_1=  \alpha_1} 
\left[ \prod_{k=2}^{n} \ket{\sigma^z_k= \alpha_{k-1} \alpha_k} \right]
\nonumber \\
\ket{\Phi_{\epsilon_{n+1} ..,\epsilon_{N-1},P }^{[n+1,..,N]\alpha_n} } && =
 \sum_{\alpha_{n+1}=\pm} ...  \sum_{\alpha_{N-1}=\pm} 
\left[ \prod_{k=n+1}^{N-1}\lambda^{[\epsilon_k] \alpha_k}_k  \right]
\left[ \prod_{k=n+1}^{N-1} \ket{\sigma^z_k= \alpha_{k-1} \alpha_k} \right]
\ket{\sigma^z_N=  \alpha_{N-1} P} 
\label{schmidtLR}
\end{eqnarray}
are the corresponding orthonormalized Schmidt eigenvectors of the Left part $[1,..,n]$
and of the Right part $[n+1,...,N]$
\begin{eqnarray}
 \braket{ \Phi_{\epsilon_1,...,\epsilon_{n-1}  }^{[1,..,n]\alpha_n'}  \vert \Phi_{\epsilon_1,...,\epsilon_{n-1}  }^{[1,..,n]\alpha_n}   } && = \delta_{\alpha_n',\alpha_n}
\nonumber \\
\braket{ \Phi_{\epsilon_{n+1} ..,\epsilon_{N-1},P }^{[n+1,..,N]\alpha_n'} \vert \Phi_{\epsilon_{n+1} ..,\epsilon_{N-1},P }^{[n+1,..,N]\alpha_n} } && = \delta_{\alpha_n',\alpha_n}
\label{schmidortho}
\end{eqnarray}

The Schmidt decomposition of Eq. \ref{mpsschmidt}
for the MPS across the bond $[k,k+1]$ 
translates into the following decomposition for the corresponding projector
\begin{eqnarray}
\ket{\psi_{\epsilon_1,...,\epsilon_{N-1},P  } } 
\bra{\psi_{\epsilon_1,...,\epsilon_{N-1},P  } } 
&& = \sum_{ \substack{\alpha_n =\pm \\ \alpha_n'=\pm} }
\left(\ket{\Phi_{\epsilon_1,...,\epsilon_{n-1}  }^{[1,..,n] \alpha_n} }
\bra{\Phi_{\epsilon_1,...,\epsilon_{n-1}  }^{[1,..,n] \alpha_n'} }
\right)
\lambda^{[\epsilon_n] \alpha_n}_n \lambda^{[\epsilon_n] \alpha_n'}_n 
\left( \ket{\Phi_{\epsilon_{n+1} ..,\epsilon_{N-1},P }^{[n+1,..,N]\alpha_n} }
\bra{\Phi_{\epsilon_{n+1} ..,\epsilon_{N-1},P }^{[n+1,..,N]\alpha_n'} }
\right)
\label{rho1bond}
\end{eqnarray}
The orthonormalization of Eqs \ref{schmidortho} yields that the trace over the Right part $[n+1,..,N] $
and the trace over the Left part $[1,..,n] $ are diagonal
\begin{eqnarray}
{\rm Tr}_{\{n+1,...,N\}} \left( \ket{\psi_{\epsilon_1,...,\epsilon_{N-1} ,P } } 
\bra{\psi_{\epsilon_1,...,\epsilon_{N-1} ,P } } \right)
&& = \sum_{ \alpha_n=\pm  }p^{[\epsilon_n] \alpha_n}_n 
\left(\ket{\Phi_{\epsilon_1,...,\epsilon_{n-1}  }^{[1,..,n] \alpha_n} }
\bra{\Phi_{\epsilon_1,...,\epsilon_{n-1}  }^{[1,..,n] \alpha_n} }
\right)
\nonumber \\
{\rm Tr}_{\{1,...,n\}} \left( \ket{\psi_{\epsilon_1,...,\epsilon_{N-1},P  } } 
\bra{\psi_{\epsilon_1,...,\epsilon_{N-1},P  } } \right)
&& = \sum_{ \alpha_n =\pm }p^{[\epsilon_n] \alpha_n}_n
\left( \ket{\Phi_{\epsilon_{n+1} ..,\epsilon_{N-1},P }^{[n+1,..,N]\alpha_n} }
\bra{\Phi_{\epsilon_{n+1} ..,\epsilon_{N-1},P }^{[n+1,..,N]\alpha_n} }
\right)
\label{rho1bonddiag}
\end{eqnarray}
with the two common weights  labelled by $\alpha_n=\pm 1$ (Eq \ref{lambdacomplexksum})
\begin{eqnarray}
p^{[\epsilon_n] \alpha_n}_n && = \left( \lambda^{[\epsilon_n] \alpha_n}_n \right)^2
=\cos^2 \left( \frac{\theta_n}{2} \right)  \delta_{\epsilon_n,\alpha_n}
+ \sin^2 \left( \frac{\theta_n}{2} \right)  \delta_{\epsilon_n,-\alpha_n}
= \frac{1+ \epsilon_n\alpha_n \cos(\theta_n) }{2}
\label{pweightsn}
\end{eqnarray}
So the corresponding entanglement entropy 
between the Left part $[1,..,n]$ and the Right part $[n+1,..,N]$
is the same for all the $2^N$ MPS of the basis and only depends on the angle $\theta_n$
associated to the bond $(n,n+1)$
\begin{eqnarray}
S^{Ent}_{([1,..n],[n+1...,N])}  && \equiv  -\sum_{\alpha_n=\pm} p^{[\epsilon_n] \alpha_n}_n \ln ( p^{[\epsilon_n] \alpha_n}_n ) 
\nonumber \\
&& 
= - \frac{1+ \cos(\theta_n) }{2} \ln \left(  \frac{1+ \cos(\theta_n) }{2} \right)
- \frac{1- \cos(\theta_n) }{2} \ln \left(  \frac{1- \cos(\theta_n) }{2} \right)
\label{entanglemententropy}
\end{eqnarray}

\subsection{ Entanglement properties of single spins in individual MPS}

For the special case $n=1$, the Left Schmidt eigenvector of Eq. \ref{schmidtLR}
reduces to
\begin{eqnarray}
\ket{\Phi^{[1] \alpha_1} }  = \ket{ \sigma_1^z=\alpha_1 } 
\label{psi1}
\end{eqnarray}
and Eq. \ref{rho1bonddiag} corresponds to the reduced density matrix for the spin $n=1$ alone in the MPS-projector
\begin{eqnarray}
&& {\rm Tr}_{\{2,...,N\}} \left( \ket{\psi_{\epsilon_1,...,\epsilon_{N-1},P  } } 
\bra{\psi_{\epsilon_1,...,\epsilon_{N-1},P  } } \right)
 = \sum_{ \alpha_1=\pm  }p^{[\epsilon_1] \alpha_1}_1 
\left(\ket{ \sigma_1^z=\alpha_1 }\bra{ \sigma_1^z=\alpha_1 }\right)
\nonumber \\
&&= \sum_{ \alpha_1=\pm  } 
\left( \frac{1+ \epsilon_1\alpha_1 \cos(\theta_1) }{2} \right)
\left( \frac{1+ \alpha_1 \sigma_1^z }{2}\right)
 = \frac{1+ \epsilon_1 \cos(\theta_1) \sigma_1^z  }{2} 
\label{rho1boundary}
\end{eqnarray}
with the corresponding entanglement entropy of Eq. \ref{entanglemententropy} that depends only on the angle $\theta_1$
\begin{eqnarray}
S^{Ent}_{([1],[2,...,N])}  && 
= - \frac{1+ \cos(\theta_1) }{2} \ln \left(  \frac{1+ \cos(\theta_1) }{2} \right)
- \frac{1- \cos(\theta_1) }{2} \ln \left(  \frac{1- \cos(\theta_1) }{2} \right)
\label{entanglemententropy1}
\end{eqnarray}

Similarly for the special case $n=N-1$, the Right Schmidt eigenvector of Eq. \ref{schmidtLR}
reduces to
\begin{eqnarray}
\ket{\Phi_{P }^{[N]\alpha_{N-1}} } && =
\ket{\sigma^z_N=  \alpha_{N-1} P} 
\label{psin}
\end{eqnarray}
and Eq. \ref{rho1bonddiag} corresponds to the reduced density matrix for the spin $n=N$ alone in the MPS-projector
\begin{eqnarray}
&& {\rm Tr}_{\{1,...,N-1\}} \left( \ket{\psi_{\epsilon_1,...,\epsilon_{N-1},P  } } 
\bra{\psi_{\epsilon_1,...,\epsilon_{N-1},P  } } \right)
 = \sum_{ \alpha_{N-1} =\pm } p^{[\epsilon_{N-1} ] \alpha_{N-1}}_{N-1}
\left( \ket{\sigma^z_N=  \alpha_{N-1} P }
\bra{ \sigma^z_N=  \alpha_{N-1} P  }
\right)
\nonumber \\
&& =  \sum_{ \alpha_{N-1} =\pm } 
\left( \frac{1+ \epsilon_{N-1}\alpha_{N-1} \cos(\theta_{N-1}) }{2} \right)
\left( \frac{1+ \alpha_{N-1} P  \sigma_N^z }{2}\right)
=  \frac{1+ \epsilon_{N-1} P  \cos(\theta_{N-1} ) \sigma_N^z }{2} 
\label{rhonbonddiag}
\end{eqnarray}
with the corresponding entanglement entropy of Eq. \ref{entanglemententropy} that depends only on the angle $\theta_{N-1}$
\begin{eqnarray}
S^{Ent}_{([1,..N-1],[N])}  && 
= - \frac{1+ \cos(\theta_{N-1}) }{2} \ln \left(  \frac{1+ \cos(\theta_{N-1}) }{2} \right)
- \frac{1- \cos(\theta_{N-1}) }{2} \ln \left(  \frac{1- \cos(\theta_{N-1}) }{2} \right)
\label{entanglemententropyn}
\end{eqnarray}

For a site $n$ in the bulk ($2 \leq n \leq n-1$), the 
simultaneous Schmidt decomposition with respect to the Left part $[1,...,n-1]$
and to the Right part $[n+1,...,N]$ reads for the MPS 
\begin{eqnarray}
\ket{\psi_{\epsilon_1,...,\epsilon_{N-1} ,P } } && = 
 \sum_{\alpha_{n-1}=\pm}
\sum_{\alpha_{n}=\pm}
\ket{\Phi_{\epsilon_1,...,\epsilon_{n-2}  }^{[1,..,n-1] \alpha_{n-1}} }
\lambda^{[\epsilon_{n-1}] \alpha_{n-1}}_{n-1}
\ket{\sigma^z_n= \alpha_{n-1} \alpha_n} 
\lambda^{[\epsilon_n] \alpha_n}_n 
\ket{\Phi_{\epsilon_{n+1} ..,\epsilon_{N-1},P }^{[n+1,..,N]\alpha_n} }
\label{mpsschmidt2}
\end{eqnarray}
and for the corresponding MPS-projector
\begin{eqnarray}
\ket{\psi_{\epsilon_1,...,\epsilon_{N-1},P  } } 
\bra{\psi_{\epsilon_1,...,\epsilon_{N-1},P  } } 
&& =  \sum_{ \substack{\alpha_{n-1} =\pm \\ \alpha_{n-1}'=\pm} }
 \sum_{ \substack{\alpha_n =\pm \\ \alpha_n'=\pm} }
\left( \ket{\Phi_{\epsilon_1,...,\epsilon_{n-2}  }^{[1,..,n-1] \alpha_{n-1}} }
\bra{\Phi_{\epsilon_1,...,\epsilon_{n-2}  }^{[1,..,n-1] \alpha_{n-1}'} }
\right)
\lambda^{[\epsilon_{n-1}] \alpha_{n-1}}_{n-1}
\lambda^{[\epsilon_{n-1}] \alpha_{n-1}'}_{n-1}
\nonumber \\
&& \left( \ket{\sigma^z_n= \alpha_{n-1} \alpha_n} 
\bra{\sigma^z_n= \alpha_{n-1}' \alpha_n'} 
\right)
\lambda^{[\epsilon_n] \alpha_n}_n 
\lambda^{[\epsilon_n] \alpha_n'}_n 
\left( \ket{\Phi_{\epsilon_{n+1} ..,\epsilon_{N-1},P }^{[n+1,..,N]\alpha_n} }
 \bra{\Phi_{\epsilon_{n+1} ..,\epsilon_{N-1},P }^{[n+1,..,N]\alpha_n'} }
\right)
\label{mposchmidt2}
\end{eqnarray}
The orthonormalization of the Schmidt eigenvectors (Eq. \ref{schmidortho}) yields
 that the trace over all the spins except $n$ 
only involves the weights of Eq. \ref{pweightsn}
corresponding to the two neighboring bonds
\begin{eqnarray}
&& {\rm Tr}_{\{1,..n-1,n+1,.,N\}} \left( \ket{\psi_{\epsilon_1,...,\epsilon_{N-1},P  } } 
\bra{\psi_{\epsilon_1,...,\epsilon_{N-1},P  } } \right)
 =  \sum_{ \alpha_{n-1} =\pm  }
 \sum_{ \alpha_n =\pm  }
p^{[\epsilon_{n-1}] \alpha_{n-1}}_{n-1}
 \left( \ket{\sigma^z_n= \alpha_{n-1} \alpha_n} 
\bra{\sigma^z_n= \alpha_{n-1} \alpha_n} 
\right)
p^{[\epsilon_n] \alpha_n}_n
\nonumber \\
&& =  \sum_{ \alpha_{n-1} =\pm  }
 \sum_{ \alpha_n =\pm  }
\left( \frac{1+ \epsilon_{n-1}\alpha_{n-1} \cos(\theta_{n-1}) }{2}\right)
\left( \frac{1+\alpha_{n-1}   \alpha_n \sigma_n^z }{2}\right)
\left( \frac{1+ \epsilon_n\alpha_n \cos(\theta_n) }{2}\right)
\nonumber \\
&& = \frac{1+ \epsilon_{n-1} \epsilon_n \cos(\theta_{n-1}) \cos(\theta_n) \sigma_n^z  }{2} 
\label{mpsschmidtrhon}
\end{eqnarray}
So the corresponding entanglement entropy 
between the site $[n]$ and its environment $[1,..,n-1,n+1,..,N]$
is the same for all the $2^N$ MPS of the basis and only depends on the angles 
$(\theta_{n-1},\theta_n)$ of the two neighboring bonds
\begin{eqnarray}
S^{Ent}_{([n],[1,..,n-1,n+1...,N])}  && 
= - \frac{1+ \cos(\theta_{n-1})\cos(\theta_n) }{2} \ln \left(  \frac{1+ \cos(\theta_{n-1})\cos(\theta_n) }{2} \right)
\nonumber \\
&& - \frac{1- \cos(\theta_{n-1})\cos(\theta_n) }{2} \ln \left(  \frac{1- \cos(\theta_{n-1})\cos(\theta_n) }{2} \right)
\label{entanglemententropysiten}
\end{eqnarray}


\subsection{ Multifractal properties of individual MPS with respect to the initial spin basis}

Even without disorder, the groundstate wavefunction of manybody quantum systems 
has been found to be generically multifractal, with many studies
concerning the Shannon-R\'enyi entropies in quantum spin models
\cite{jms2009,jms2010,jms2011,moore,grassberger,atas_short,
atas_long,luitz_short,luitz_o3,luitz_spectro,luitz_qmc,jms2014,alcaraz1,alcaraz2,c_renyi,jms2017,c_treetensor},
while multifractal properties have been also studied recently for excited states in the field of Many-Body-Localization
\cite{alet,mace1,mace2,lev}.
In our present framework,
 the expansion of the MPS of Eq \ref{mpsfinal}
in the Pauli basis $\sigma_k^z=\pm 1 $
\begin{eqnarray}
\ket{\psi_{\epsilon_1,...,\epsilon_{N-1},P  } } 
= \sum_{S_1=\pm 1}\sum_{S_2=\pm 1} ... \sum_{S_N=\pm 1} 
\psi^{S_1,...,S_N}_{\epsilon_1,...,\epsilon_{N-1},P  }
\ket {\sigma^z_1=S_1 }\ket {\sigma^z_2=S_2 } ... \ket {\sigma^z_N=S_N }
\label{kettau}
\end{eqnarray}
involves the $2^N$ coefficients $\psi^{S_1=\pm1 ,...,S_N=\pm 1}_{\epsilon_1,...,\epsilon_{N-1},P  } $ given by
\begin{eqnarray}
\psi^{S_1,...,S_N}_{\epsilon_1,...,\epsilon_{N-1},P  }
 && = \braket{ S_1,...,S_N \vert \psi_{\epsilon_1,...,\epsilon_{N-1},P  }}
=  \sum_{\alpha_1=\pm} ...  \sum_{\alpha_{N-1}=\pm} 
\left[ \prod_{k=1}^{N-1}\lambda^{[\epsilon_k] \alpha_k}_k  \right]
\delta_{S_1,\alpha_1} 
\left[ \prod_{k=2}^{N-1} \delta_{S_k, \alpha_{k-1} \alpha_k} \right]
\delta_{S_N,  \alpha_{N-1} P} 
\nonumber \\
&& = \delta_{P, \prod_{j=1}^N S_j} \prod_{k=1}^{N-1}\lambda^{[\epsilon_k] \prod_{n=1}^k S_n }_k 
\label{mpscoef}
\end{eqnarray}
The statistics of the corresponding $2^N$ weights $\vert \psi^{S_1=\pm1 ,...,S_N=\pm 1}_{\epsilon_1,...,\epsilon_{N-1},P  } \vert^2 $
 normalized to unity can be analyzed via the Inverse Participation Ratios 
 where $q$ is a continuous parameter 
\begin{eqnarray}
Y^{(q)}_{\epsilon_1,...,\epsilon_{N-1},P  }  && \equiv  \sum_{S_1=\pm 1}\sum_{S_2=\pm 1} ... \sum_{S_N=\pm 1} 
\vert \psi^{S_1,...,S_N}_{\epsilon_1,...,\epsilon_{N-1},P  } \vert^{2q}
 = \sum_{\alpha_1=\pm 1}\sum_{\alpha_2=\pm 1} ... \sum_{\alpha_{N-1}=\pm 1} 
\prod_{k=1}^{N-1} \left[\lambda^{[\epsilon_k] \alpha_k}_k  \right]^{2q}
\nonumber \\
&&= \prod_{k=1}^{N-1} \left(  \left[\lambda^{[\epsilon_k]+}_{k} \right]^{2q}
+\left[\lambda^{[\epsilon_k]-}_{k} \right]^{2q}
\right)
= \prod_{k=1}^{N-1} \left(  \cos^{2q} \left( \frac{\theta_{k}}{2} \right)
+\sin^{2q} \left( \frac{\theta_{k}}{2} \right)
\right)
\label{yq}
\end{eqnarray}
that are actually independent of the precise MPS $\ket{\psi_{\epsilon_1,...,\epsilon_{N-1},P  } } $.
So the $2^N$ MPS of the basis have all the same R\'enyi entropy
 \begin{eqnarray}
{\cal S}_q(N) \equiv  \frac{ \ln Y^{(q)}_{\epsilon_1,...,\epsilon_{N-1},P  } }{1-q} 
= \frac{1}{1-q} \sum_{k=1}^{N-1} \ln \left(  
\cos^{2q} \left( \frac{\theta_{k}}{2} \right)
+\sin^{2q} \left( \frac{\theta_{k}}{2}\right)
\right)
\label{renyi}
\end{eqnarray}
and the same generalized fractal dimensions $0 \leq {\cal D}_q \leq 1$
that describe the leading extensive behaviors
 \begin{eqnarray}
S_q(N)  \oppropto_{N \to +\infty} {\cal D}_q  (N \ln 2)
\label{renyid}
\end{eqnarray}


\section{ Corresponding Matrix-Product-Operator basis for operators }

\label{sec_MPObasis}

\subsection{ From the MPS basis to the MPO basis  }

The MPS basis of Eq. \ref{mpsfinal} for the Hilbert space of size $2^N$
can be used to construct the following basis for the operator-space of size $4^N$
\begin{eqnarray}
\ket{\psi_{\epsilon_1,...,\epsilon_{N-1},P } } 
 \bra{\psi_{\epsilon_1',...,\epsilon_{N-1}',P' } } 
= \sum_{ \substack{\alpha_1 =\pm \\ \alpha_1'=\pm} }
...
\sum_{ \substack{\alpha_{N-1} =\pm \\ \alpha_{N-1}'=\pm} }
\left[ \prod_{k=1}^{N-1}  
\lambda^{[\epsilon_k] \alpha_k}_k 
\lambda^{[\epsilon_k'] \alpha_k'}_k  
 \right]
O_1^{  \alpha_1, \alpha_1'} 
\left[ \prod_{k=2}^{N-1} O_k^{ \alpha_{k-1} \alpha_k,  \alpha_{k-1}' \alpha_k'} 
\right]
O_N^{ \alpha_{N-1} P , \alpha_{N-1}' P' }
\label{mpoproj}
\end{eqnarray}
where the elementary operators that appear in these $4^N$ Matrix-Product-Operators
read in terms of the Pauli matrices $\sigma_k^{x,y,z}$
\begin{eqnarray}
O_k^{S,S'}  \equiv \ket{\sigma^z_k=S }\bra{\sigma^z_k=S' } 
= \delta_{S,S'} \left( \frac{1+S \sigma_k^z}{2}\right)
+ \delta_{S,-S'} \left( \frac{ \sigma^x_k +i S \sigma^y_k }{2}\right)
\label{operatorfinal}
\end{eqnarray}

\subsection{ Explicit form of the parity pseudo-spin operators $P^{x,y,z}$ }

Using the following property satisfied by the bond variables of Eq. \ref{lambdacomplexksum}
\begin{eqnarray}
\sum_{\epsilon_k=\pm }  \lambda^{[\epsilon_k] \alpha_k}_k \lambda^{[\epsilon_k] \alpha_k'}_k    
 = \delta_{\alpha_k,\alpha_k'} 
\label{lambdabondrhosumeps}
\end{eqnarray}
one can check that the sum over the $2^{N-1}$ projectors associated to the fixed parity value $P$
\begin{eqnarray}
&& \sum_{\epsilon_1=\pm } ...\sum_{\epsilon_{N-1}=\pm } 
\ket{\psi_{\epsilon_1,...,\epsilon_{N-1},P } } 
 \bra{\psi_{\epsilon_1,...,\epsilon_{N-1},P } } 
 \nonumber \\ &&
 = \sum_{ \substack{\alpha_1 =\pm \\ \alpha_1'=\pm} }
...
\sum_{ \substack{\alpha_{N-1} =\pm \\ \alpha_{N-1}'=\pm} }
\left[ \prod_{k=1}^{N-1} \left( \sum_{\epsilon_k=\pm }  
\lambda^{[\epsilon_k] \alpha_k}_k \lambda^{[\epsilon_k] \alpha_k'}_k  
\right)
 \right]
O_1^{  \alpha_1, \alpha_1'} 
\left[ \prod_{k=2}^{N-1} O_k^{ \alpha_{k-1} \alpha_k,  \alpha_{k-1}' \alpha_k'} 
\right]
 O_N^{ \alpha_{N-1} P, \alpha_{N-1}' P}
\nonumber \\
&& = \sum_{\alpha_1=\pm} ...  \sum_{\alpha_{N-1}=\pm} 
O_1^{  \alpha_1, \alpha_1} 
\left[ \prod_{k=2}^{N-1} O_k^{ \alpha_{k-1} \alpha_k,  \alpha_{k-1} \alpha_k} 
\right]
O_N^{ \alpha_{N-1} P, \alpha_{N-1} P}
\nonumber \\
&& = \sum_{\alpha_1=\pm} ...  \sum_{\alpha_{N-1}=\pm} 
\left( \frac{1+\alpha_1 \sigma_1^z}{2}\right)
\left[ \prod_{k=2}^{N-1} \left( \frac{1+ \alpha_{k-1} \alpha_k\sigma_k^z}{2}\right)
\right]
\left( \frac{1+ \alpha_{N-1} P \sigma_N^z}{2}\right) 
\nonumber \\
&& =\frac{1}{2} \left[  1+ P \prod_{k=1}^N \sigma_k^z \right] 
\equiv \frac{1}{2} \left[  1+ P P^z \right] 
\label{identityprojparity}
\end{eqnarray}
gives the projector on the eigenvalue $P$ of the parity operator $P^z$ as it should.
When one sums over the two parity values $P=\pm 1$, one obtains the decomposition of the identity 
\begin{eqnarray}
1 = &&  \sum_{\epsilon_1=\pm } ...\sum_{\epsilon_{N-1}=\pm } \sum_{P=\pm 1}
\ket{\psi_{\epsilon_1,...,\epsilon_{N-1},P } } 
 \bra{\psi_{\epsilon_1,...,\epsilon_{N-1},P } } 
\label{identityproj}
\end{eqnarray}
while the sum weighted by the parity value $P$ gives the 
decomposition of the parity operator of Eq. \ref{paritysigmaz}
\begin{eqnarray}
P^z && =  \sum_{\epsilon_1=\pm } ...\sum_{\epsilon_{N-1}=\pm } 
\sum_{P=\pm 1}  P \ket{\psi_{\epsilon_1,...,\epsilon_{N-1},P } } 
 \bra{\psi_{\epsilon_1,...,\epsilon_{N-1},P } }  
\label{pzproj}
\end{eqnarray}
For given values of the $(N-1)$ bond variables $(\epsilon_1,..,\epsilon_{N-1})$,
the two MPS of Eq. \ref{mpsfinal} associated to the two possible parity eigenvalues $P=\pm 1$
are simply related by the flip operator $\sigma_N^x$ of the last spin $N$
\begin{eqnarray}
\sigma_N^x \ket{\psi_{\epsilon_1,...,\epsilon_{N-1} ,P } } && = \ket{\psi_{\epsilon_1,...,\epsilon_{N-1},-P  } } 
\label{pairingketp}
\end{eqnarray}
so the flip operator $P^x$ of the parity reduces to this boundary flip operator $\sigma_N^x$
\begin{eqnarray}
P^x && 
\equiv \sum_{\epsilon_1=\pm } ...\sum_{\epsilon_{N-1}=\pm } 
\sum_{P=\pm 1} \ket{\psi_{\epsilon_1,...,\epsilon_{N-1},P } } 
 \bra{\psi_{\epsilon_1,...,\epsilon_{N-1},-P } } =\sigma_N^x
\label{pxparityflip}
\end{eqnarray}
The third operator $P^y$ of the pseudospin associated to the parity can be then obtained
via
\begin{eqnarray}
P^y && = -i P^z P^x =   \left( \prod_{j=1}^{N-1} \sigma_j^z \right) \sigma_N^y
\label{pyproj}
\end{eqnarray}

\subsection{ Definition of the pseudo-spins operators $\epsilon_k^{x,y,z} $ in the MPS basis and in the MPO basis}

The pseudo-spins operators $(\epsilon^z_n,\epsilon_n^x)$ associated to the labels $\epsilon_n=\pm 1$ of the MPS of Eq. \ref{mpsfinal}
can be defined by their actions in the MPS basis :
the operator $\epsilon_n^z$ reads the value $\epsilon_n$ of each MPS
\begin{eqnarray}
\epsilon^z_n  \ket{\psi_{\epsilon_1,...,\epsilon_{N-1},P } } && \equiv \epsilon_n  \ket{\psi_{\epsilon_1,...,\epsilon_{N-1},P } }
\label{enzdef}
\end{eqnarray}
while the operator $\epsilon_n^x$ flips the value $\epsilon_n$ of each MPS
\begin{eqnarray}
\epsilon^x_n  \ket{\psi_{\epsilon_1,..,\epsilon_n,..,\epsilon_{N-1},P }  } && \equiv \ket{\psi_{\epsilon_1,.,-\epsilon_n,..,\epsilon_{N-1},P } }
\label{enxdef}
\end{eqnarray}
As a consequence, $\epsilon^z_n$ and $\epsilon^x_n$ anticommute,
the third pseudo-spin operator can be defined via
\begin{eqnarray}
\epsilon^y_n  && = -i \epsilon^z_n \epsilon^x_n 
\label{enydef}
\end{eqnarray}
and the pseudo-spin operators associated to different sites $n \ne m$ commute.

Their actions in the full Hilbert space can be then obtained
from their expansions in the MPO basis of Eq. \ref{mpoproj}
\begin{eqnarray}
\epsilon^z_n && \equiv \sum_{P=\pm 1} \sum_{\epsilon_1=\pm } ...\sum_{\epsilon_{N-1}=\pm } 
\epsilon_n   \ket{\psi_{\epsilon_1,....,\epsilon_{N-1},P } } 
 \bra{\psi_{\epsilon_1,...,\epsilon_{N-1},P } } 
\nonumber \\
\epsilon^x_n && \equiv \sum_{P=\pm 1} \sum_{\epsilon_1=\pm } ...\sum_{\epsilon_{N-1}=\pm } 
\ket{\psi_{\epsilon_1,..,\epsilon_n,..,\epsilon_{N-1},P } } 
 \bra{\psi_{\epsilon_1,.,-\epsilon_n,..,\epsilon_{N-1},P } } 
\label{enzxhilbert}
\end{eqnarray}

\subsection{ Explicit form of the pseudo-spin operators $\epsilon_n^{x,y,z}$ in terms of the initial spin operators $\sigma_k^{x,y,z} $}

Since the operators $\epsilon^{z,x}_n$ act only locally on the bond variable $\epsilon_n$
in the MPO basis of Eq. \ref{mpoproj}, one can rewrite Eqs \ref{enzxhilbert} as
\begin{eqnarray}
\epsilon_n^z && = \sum_{\alpha_n=\pm}  \sum_{\alpha_n'=\pm} 
L_{[1,..,n]}^{\alpha_{n} ,\alpha_n'} \ 
Z_{n,n+1}^{\alpha_n,\alpha_n'} \ 
R_{[n+1,..,N]}^{\alpha_{n} ,\alpha_n'} 
\nonumber \\
\epsilon^x_n && 
=   \sum_{\alpha_n=\pm}  \sum_{\alpha_n'=\pm} 
L_{[1,..,n]}^{\alpha_{n} ,\alpha_n'} \ 
X_{n,n+1}^{\alpha_n,\alpha_n'} \ 
R_{[n+1,..,N]}^{\alpha_{n} ,\alpha_n'} 
\label{enzxLR}
\end{eqnarray}
where the central terms take into account the specific actions on the binary variable $\epsilon_n=\pm 1$
of the bond $(n,n+1)$
\begin{eqnarray}
Z_n^{\alpha_n,\alpha_n'} && \equiv \sum_{\epsilon_n=\pm }  \epsilon_n
\lambda^{[\epsilon_n] \alpha_n}_n  \lambda^{[\epsilon_n] \alpha_n'}_n
 = \delta_{\alpha_n,\alpha_n'} \alpha_n \cos (\theta_n) 
+  \delta_{\alpha_n,- \alpha_n'} \sin (\theta_n)
\nonumber \\
X_n^{\alpha_n,\alpha_n'} && \equiv \lambda^{[+] \alpha_n}_n  \lambda^{[-] \alpha_n'}_n
+ \lambda^{[-] \alpha_n}_n  \lambda^{[+] \alpha_n'}_n 
=\delta_{\alpha_n,\alpha_n'}  \alpha_n \sin (\theta_n) 
-  \delta_{\alpha_n,- \alpha_n'} \cos (\theta_n) 
\label{enx}
\end{eqnarray}
while the Left term resums the MPO on the Left part $[1,..,n]$ 
\begin{eqnarray}
L_{[1,..,n]}^{\alpha_{n} ,\alpha_n'} && \equiv \sum_{\alpha_{n-1}=\pm} 
\left(
\sum_{\alpha_1=\pm} ...  \sum_{\alpha_{n-2}=\pm} 
O_1^{  \alpha_1, \alpha_1} 
\left[ \prod_{k=2}^{n-1} O_k^{ \alpha_{k-1} \alpha_k,  \alpha_{k-1} \alpha_k} 
\right]
\right)
O_n^{ \alpha_{n-1} \alpha_n,  \alpha_{n-1} \alpha_n'} 
\nonumber \\
&& =
\sum_{\alpha_{n-1}=\pm} 
\left(\sum_{\alpha_1=\pm} ...  \sum_{\alpha_{n-2}=\pm} 
\left( \frac{1+\alpha_1 \sigma_1^z}{2}\right)
\left[ \prod_{k=2}^{n-1} 
\left( \frac{1+\alpha_{k-1} \alpha_k \sigma_k^z}{2}\right)
\right] 
\right)
O_n^{ \alpha_{n-1} \alpha_n,  \alpha_{n-1} \alpha_n'} 
\nonumber \\
&& =
\frac{1}{4} \sum_{\alpha_{n-1}=\pm} 
 \left[  1+\alpha_{n-1}  \prod_{k=1}^{n-1} \sigma_k^z \right]
\left[ \delta_{\alpha_n,\alpha_n'} \left( 1+\alpha_{n-1} \alpha_n \sigma_n^z\right)
+ \delta_{\alpha_n,-\alpha_n'} \left(  \sigma^x_n +i \alpha_{n-1} \alpha_n \sigma^y_n \right)
\right]
\nonumber \\
&& =\frac{1}{2} 
\left[ \delta_{\alpha_n,\alpha_n'} \left( 1+ \alpha_n \prod_{k=1}^{n} \sigma_k^z   \right)
+ \delta_{\alpha_n,-\alpha_n'} \left(  \sigma^x_n +i  \alpha_n \left( \prod_{k=1}^{n-1} \sigma_k^z\right)\sigma^y_n \right)
\right]
\label{leftresum}
\end{eqnarray}
and the Right term resums the MPO on the Right part $[n+1,N]$
\begin{eqnarray}
R_{[n+1,..,N]}^{\alpha_{n} ,\alpha_n'}  && \equiv 
\sum_{\alpha_{n+1}=\pm}O_{n+1}^{ \alpha_{n} \alpha_{n+1},  \alpha_n' \alpha_{n+1}} 
\left(
 \sum_{\alpha_{n+2}=\pm} ...  \sum_{\alpha_{N-1}=\pm} 
\left[ \prod_{k=n+2}^{N-1} O_k^{ \alpha_{k-1} \alpha_k,  \alpha_{k-1} \alpha_k} 
\right]
\sum_{P=\pm 1} O_N^{ \alpha_{N-1} P, \alpha_{N-1} P}
\right)
\nonumber \\
&&
=\sum_{\alpha_{n+1}=\pm}O_{n+1}^{ \alpha_{n} \alpha_{n+1},  \alpha_n' \alpha_{n+1}} 
\left(
  \sum_{\alpha_{n+2}=\pm} ...  \sum_{\alpha_{N-1}=\pm} 
\left[ \prod_{k=n+2}^{N-1} \left( \frac{1+\alpha_{k-1} \alpha_k \sigma_k^z}{2}\right)
\right]
\sum_{P=\pm 1} \left( \frac{1+ \alpha_{N-1} P  \sigma_N^z}{2}\right) 
\right)
\nonumber \\
&& =
\frac{1}{2}\sum_{\alpha_{n+1}=\pm}
\left[ \delta_{\alpha_n,\alpha_n'} \left( 1+\alpha_{n} \alpha_{n+1} \sigma_{n+1}^z\right)
+ \delta_{\alpha_n,-\alpha_n'} \left(  \sigma^x_{n+1} +i \alpha_{n} \alpha_{n+1}  \sigma^y_{n+1} \right)
\right]
\nonumber \\
&& =\delta_{\alpha_n,\alpha_n'} 
+ \delta_{\alpha_n,-\alpha_n'} \sigma^x_{n+1}
\label{rightresum}
\end{eqnarray}

Putting everything together in Eqs \ref{enzxLR},
one obtains the final results for the pseudo-spin operator $\epsilon_n^z $
\begin{eqnarray}
\epsilon_n^z
&&=\frac{1}{2}  \sum_{\alpha_n=\pm}  \sum_{\alpha_n'=\pm} 
\left[ \delta_{\alpha_n,\alpha_n'} \left( 1+ \alpha_n \prod_{k=1}^{n} \sigma_k^z   \right)
+ \delta_{\alpha_n,-\alpha_n'} \left(  \sigma^x_n +i  \alpha_n \left( \prod_{k=1}^{n-1} \sigma_k^z\right)\sigma^y_n \right)
\right]
\nonumber \\
&& 
\left[ \delta_{\alpha_n,\alpha_n'} 
  \alpha_n \cos (\theta_n) 
+  \delta_{\alpha_n,- \alpha_n'}
\sin (\theta_n) 
\right]
\left(  \delta_{\alpha_n,\alpha_n'} 
+ \delta_{\alpha_n,-\alpha_n'}  \sigma^x_{n+1} 
\right)
\nonumber \\
&&=\frac{1}{2}  \sum_{\alpha_n=\pm}  
\left[  \cos (\theta_n)  \left(  \alpha_n + \prod_{k=1}^{n} \sigma_k^z   \right)
+ 
\sin (\theta_n) \left(  \sigma^x_n +i  \alpha_n \left( \prod_{k=1}^{n-1} \sigma_k^z\right)\sigma^y_n \right)\sigma^x_{n+1} 
\right]
\nonumber \\
&&=  \cos (\theta_n) \left(   \prod_{k=1}^{n} \sigma_k^z   \right)
+\sin (\theta_n)  \sigma^x_n   \sigma^x_{n+1} 
\label{enzsigma}
\end{eqnarray}
and for the pseudo-spin operator $\epsilon_n^x $
\begin{eqnarray}
\epsilon_n^x
&&=\frac{1}{2}  \sum_{\alpha_n=\pm}  \sum_{\alpha_n'=\pm} 
\left[ \delta_{\alpha_n,\alpha_n'} \left( 1+ \alpha_n \prod_{k=1}^{n} \sigma_k^z   \right)
+ \delta_{\alpha_n,-\alpha_n'} \left(  \sigma^x_n +i  \alpha_n \left( \prod_{k=1}^{n-1} \sigma_k^z\right)\sigma^y_n \right)
\right]
\nonumber \\
&& 
\left[\delta_{\alpha_n,\alpha_n'}  \alpha_n \sin (\theta_n) 
-  \delta_{\alpha_n,- \alpha_n'} \cos (\theta_n) 
\right]
\left(  \delta_{\alpha_n,\alpha_n'} 
+ \delta_{\alpha_n,-\alpha_n'}  \sigma^x_{n+1} 
\right)
\nonumber \\
&&=\frac{1}{2}  \sum_{\alpha_n=\pm}  
\left[   \sin (\theta_n) \left(  \alpha_n + \prod_{k=1}^{n} \sigma_k^z   \right)
-  \cos (\theta_n) \left(  \sigma^x_n +i  \alpha_n \left( \prod_{k=1}^{n-1} \sigma_k^z\right)\sigma^y_n \right)\sigma^x_{n+1} 
\right]
\nonumber \\
&&=
 \sin (\theta_n) \left(   \prod_{k=1}^{n} \sigma_k^z   \right)
-  \cos (\theta_n)  \sigma^x_n \sigma^x_{n+1} 
\label{enxsigma}
\end{eqnarray}
while the third pseudo-spin operator $\epsilon_n^y $ is then obtained using Eq. \ref{enydef}
\begin{eqnarray}
\epsilon_n^y
&&=- i \epsilon_n^z \epsilon_n^x=
-i \left[ 
\cos (\theta_n) \left(   \prod_{k=1}^{n} \sigma_k^z   \right)
+\sin (\theta_n)  \sigma^x_n   \sigma^x_{n+1} 
\right]
\left[ 
 \sin (\theta_n) \left(   \prod_{k=1}^{n} \sigma_k^z   \right)
-  \cos (\theta_n)  \sigma^x_n \sigma^x_{n+1} 
\right]
\nonumber \\
&& =
 - \left(   \prod_{k=1}^{n-1} \sigma_k^z   \right) \sigma^y_n \sigma^x_{n+1} 
\label{enysigma}
\end{eqnarray}

The interpretation of these pseudo-spins in terms of Majorana fermions can be found in Appendix \ref{app_majorana}.

\subsection{ Initial spin operators $\sigma_k^{x,y,z} $
in terms of the pseudo-spin operators $\epsilon_n^{x,y,z}$ }

Since the two pseudo-spin operators $\epsilon_n^{z} $ (Eq. \ref{enzsigma})
and $\epsilon_n^{x} $ (Eq. \ref{enxsigma})
are the linear combinations of the two same operators 
$\left(   \prod_{k=1}^{n} \sigma_k^z   \right)$ and $\sigma^x_n   \sigma^x_{n+1} $
of the initial Pauli basis, one can directly invert them to obtain these operators in terms of the pseudo-spins.
The first linear combination of Eqs \ref{enzsigma} and \ref{enxsigma}
 gives for $1 \leq n \leq N-1$
\begin{eqnarray}
  \prod_{k=1}^{n} \sigma_k^z   
&&=\cos (\theta_n) \epsilon_n^z
+\sin (\theta_n) \epsilon_n^x
\label{inversionstringz}
\end{eqnarray}
The bulk operator $\sigma_n^z$ for $2 \leq n \leq N-1$ 
is thus given by the following product involving only the two neighboring bonds
\begin{eqnarray}
\sigma_n^z   && =  
\left[ \cos (\theta_{n-1}) \epsilon_{n-1}^z+\sin (\theta_{n-1}) \epsilon_{n-1}^x \right]
\left[ \cos (\theta_n) \epsilon_n^z+\sin (\theta_n) \epsilon_n^x \right]
\label{inversionz}
\end{eqnarray}
while the boundary case $n=1$ involves a single term
and the boundary case $n=N$ involves the parity operator
\begin{eqnarray}
\sigma_1^z && = \cos (\theta_1) \epsilon_1^z +\sin (\theta_1) \epsilon_1^x
\nonumber \\
\sigma_N^z   && =  \left[ \prod_{k=1}^{N-1} \sigma_k^z   \right] P^z
=  \left[ \cos (\theta_{N-1}) \epsilon_{N-1}^z+\sin (\theta_{N-1}) \epsilon_{N-1}^x \right] P^z
\label{inversionzbords}
\end{eqnarray}

The second linear combination of Eqs \ref{enzsigma}
and \ref{enxsigma}
directly gives for $1 \leq n \leq N-1$
\begin{eqnarray}
\sigma^x_n \sigma^x_{n+1} 
&& =  \sin (\theta_n) \epsilon_n^z
-\cos (\theta_n) \epsilon_n^x
\label{inversionxx}
\end{eqnarray}
The product of this equation for $n=k,..,N-1$ then gives
\begin{eqnarray}
\sigma_k^x \sigma_N^x = \prod_{n=k}^{N-1} \sigma^x_n \sigma^x_{n+1} 
&& = \prod_{n=k}^{N-1} \left[ \sin (\theta_n) \epsilon_n^z
-\cos (\theta_n) \epsilon_n^x \right]
\label{inversionxxprod}
\end{eqnarray}
Since $\sigma_N^x$ corresponds to the parity flip operator $P^x$ (Eq. \ref{pxparityflip}),
one obtains that $\sigma_k^x$ involves a string of factors up to the right boundary $N$
\begin{eqnarray}
\sigma_k^x 
&& = \left(\prod_{n=k}^{N-1} \left[ \sin (\theta_n) \epsilon_n^z
-\cos (\theta_n) \epsilon_n^x \right] \right) P^x
\label{inversionx}
\end{eqnarray}
Finally, the third spin operator can be obtained from Eqs \ref{inversionz}
and \ref{inversionx}
\begin{eqnarray}
\sigma_n^y   && =  -i \sigma_n^z \sigma_n^x
= -
\left[ \cos (\theta_{n-1}) \epsilon_{n-1}^z+\sin (\theta_{n-1}) \epsilon_{n-1}^x \right]
\epsilon_n^y
\left(\prod_{k=n+1}^{N-1} \left[ \sin (\theta_k) \epsilon_k^z
-\cos (\theta_k) \epsilon_k^x \right] \right) P^x
\label{inversiony}
\end{eqnarray}
while the boundary cases $n=1$ and $n=N$ read using Eq \ref{inversionzbords}
\begin{eqnarray}
\sigma_1^y = -i \sigma_1^z \sigma_1^x && =
- \epsilon_1^y
\left(\prod_{n=2}^{N-1} \left[ \sin (\theta_n) \epsilon_n^z
-\cos (\theta_n) \epsilon_n^x \right] \right) P^x
\nonumber \\
\sigma_N^y
=-i \sigma_N^z  \sigma_N^x && 
=   \left[ \cos (\theta_{N-1}) \epsilon_{N-1}^z+\sin (\theta_{N-1}) \epsilon_{N-1}^x \right] P^y
\label{inversionybords}
\end{eqnarray}

From these single operators, one can translate any operator
written in the Pauli basis of the initial spins $(\sigma_{k=1,..,N}^{a=0,x,y,z} )$
into its form in the Pauli basis of the pseudo-spins $(\epsilon_{k=1,..,N-1}^{a=0,x,y,z} ) $ and of the parity operators $P^{0,x,y,z}$.
For instance, the simplest operators involving two neighboring spins 
in the bulk read
\begin{eqnarray}
\sigma_n^z  \sigma_{n+1}^z  && =  
\left[ \cos (\theta_{n-1}) \epsilon_{n-1}^z+\sin (\theta_{n-1}) \epsilon_{n-1}^x \right]
\left[ \cos (\theta_{n+1}) \epsilon_{n+1}^z+\sin (\theta_{n+1}) \epsilon_{n+1}^x \right]
\nonumber \\
\sigma_n^y  \sigma_{n+1}^y  && =  
\left[ \cos (\theta_{n-1}) \epsilon_{n-1}^z+\sin (\theta_{n-1}) \epsilon_{n-1}^x \right]
\left[ \sin (\theta_n) \epsilon_n^z
-\cos (\theta_n) \epsilon_n^x \right]
\left[ \cos (\theta_{n+1}) \epsilon_{n+1}^z+\sin (\theta_{n+1}) \epsilon_{n+1}^x \right]
\label{inversionzz}
\end{eqnarray}
while $\sigma^x_n \sigma^x_{n+1}  $ has been already given in Eq. \ref{inversionxx}.


\section{ Construction of Parent Hamiltonians that have these MPS as eigenstates }

\label{sec_Hamilton}

\subsection{ Parametrization of the $2^N$ energy levels }

To simplify the notations in this section, it is 
convenient to relabel the parity operators $P^{a=x,y,z}$ as
\begin{eqnarray}
\epsilon_N^a \equiv P^a
\label{epsN}
\end{eqnarray}
so that the $2^N$ MPS kets of Eq. \ref{mpsfinal}
are now parametrized by the $N$ binary variable $\epsilon_k=\pm 1$ for $k=1,..,N$.
We wish to construct the parent Hamiltonians that have these $2^N$ MPS kets as eigenvectors
\begin{eqnarray}
H = && \sum_{\epsilon_1=\pm } ...\sum_{\epsilon_{N}=\pm } 
E_{\epsilon_1,...,\epsilon_{N}  } \ket{\psi_{\epsilon_1,...,\epsilon_{N-1},\epsilon_{N} } } 
 \bra{\psi_{\epsilon_1,...,\epsilon_{N-1},\epsilon_{N} } } 
\label{hamiltonian}
\end{eqnarray}
where the $2^{N}$ energies can be parametrized in terms of $2^N$ 
couplings $J_{n_1,n_2,..,n_p}^{(p)} $ with $p=0,1,..,N$ and $1 \leq n_1 < n_2 <...< n_p  \leq N $
\begin{eqnarray}
E_{\epsilon_1,...,\epsilon_{N}  }
&& = \sum_{p=0}^N \sum_{1 \leq n_1 < n_2 <...< n_p  \leq N}  J_{n_1,n_2,..,n_p}^{(p)} \epsilon_{n_1}\epsilon_{n_2} ... \epsilon_{n_p} 
\nonumber \\
&& =
 J^{(0)} + \sum_{n=1}^{N} J_n^{(1)} \epsilon_n
+ \sum_{1 \leq n_1 < n_2 \leq N}  J_{n_1,n_2}^{(2)} \epsilon_{n_1}\epsilon_{n_2}
+ ...
+ J^{(N)} \prod_{k=1}^{N} \epsilon_k
\label{energylevela}
\end{eqnarray}
The coupling $ J^{(0)} $ corresponds to the average energy over the $2^N$ levels
 and can be chosen to vanish
\begin{eqnarray}
 J^{(0)} = \sum_{\epsilon_1=\pm} ... \sum_{\epsilon_N=\pm}  E_{\epsilon_1,...,\epsilon_{N}  } =0
\label{energyav0}
\end{eqnarray}
Since the variance of the energy over the spectrum should be extensive in $N$,
 the following rescaled variance should remain finite in the thermodynamic limit $N \to +\infty$
\begin{eqnarray}
v_N^2 && \equiv  \frac{1}{N} \sum_{\epsilon_1=\pm} ... \sum_{\epsilon_N=\pm}  E^2_{\epsilon_1,...,\epsilon_{N}  } 
= \frac{1}{N}\sum_{p=0}^N \sum_{1 \leq n_1 < n_2 <...< n_p  \leq N} 
 \left[ J_{n_1,n_2,..,n_p}^{(p)}\right]^2
\nonumber \\
&& =  \frac{1}{N}\sum_{n=1}^{N} \left[J_n^{(1)} \right]^2
+ \frac{1}{N} \sum_{1 \leq n_1 < n_2 \leq N} \left[ J_{n_1,n_2}^{(2)} \right]^2
+\frac{1}{N} \sum_{1 \leq n_1 < n_2<n_3 \leq N} \left[ J_{n_1,n_2,n_3}^{(3)} \right]^2
+ ...
+ \frac{1}{N} \left[J^{(N)} \right]^2
\label{energyvar}
\end{eqnarray}
so the couplings $J_{n_1,n_2,..,n_p}^{(p)}  $ should be chosen to
decay sufficiently rapidly as a function of
the distance along the chain.

\subsection{ Parent Hamiltonians in terms of the Local Integrals of Motion $\epsilon_n^z$ (LIOMs)}

Since $(\epsilon_1,...,\epsilon_N)$ are the eigenvalues of the commuting operators $(\epsilon_1^z,...,\epsilon_N^z)$ studied in the previous section,
Eq. \ref{energylevela} can be directly translated 
at the operator level for the Hamiltonian of Eq. \ref{hamiltonian} as
\begin{eqnarray}
H
&& =\sum_{p=1}^N \sum_{1 \leq n_1 < n_2 <...< n_p  \leq N}  J_{n_1,n_2,..,n_p}^{(p)} \epsilon^z_{n_1}\epsilon^z_{n_2} ... \epsilon^z_{n_p} 
\nonumber \\
&& =
 \sum_{n=1}^{N} J_n^{(1)} \epsilon^z_n
+ \sum_{1 \leq n_1 < n_2 \leq N}  J_{n_1,n_2}^{(2)} \epsilon^z_{n_1}\epsilon^z_{n_2}
+ ...
+ J^{(N)}_{1,2,..,N} \prod_{k=1}^{N} \epsilon^z_k
\label{hlioms}
\end{eqnarray}
where the pseudo-spin operators $\epsilon_n^z$ are called the Local Integrals of Motion (LIOMs)
in the field of Many-Body-Localization.
The first contribution 
\begin{eqnarray}
H^{(1)} \equiv \sum_{n=1}^{N} J_n^{(1)} \epsilon^z_n
=\sum_{n=1}^{N} J_n^{(1)} \left[\cos (\theta_n) \left(   \prod_{k=1}^{n} \sigma_k^z   \right)
+\sin (\theta_n)  \sigma^x_n   \sigma^x_{n+1} 
\right]
\label{freefermions}
\end{eqnarray}
is non-interacting for the LIOMS $\epsilon_n^z$ (free-fermions) 
and the corresponding couplings $ J_n^{(1)}$ can be chosen to be random of order $O(1)$.
The other terms $2 \leq p \leq N$ correspond to the most general interactions between the LIOMS $\epsilon_n^z$,
where the couplings have to satisfy the extensivity constraint of Eq. \ref{energyvar}.
However, if one wishes to construct local Hamiltonians, one can choose to keep 
only the interactions between the nearest-neighbor LIOMS $\epsilon^z_n $ and $\epsilon^z_{n+1} $ 
(already translated in terms of the initial spin operators in Eq. \ref{inversionzz})
\begin{eqnarray}
&& H^{(2)nn} \equiv \sum_{n=1}^{N-1} J_{n,n+1}^{(2)} \epsilon^z_n \epsilon^z_{n+1}
=  \sum_{n=1}^{N-1} J_{n,n+1}^{(2)}
\label{h2nn}
\nonumber \\ &&
\left[\cos (\theta_n)  \cos (\theta_{n+1}) \sigma^z_{n+1} 
+\sin (\theta_n) \sin (\theta_{n+1})\sigma^x_n    \sigma^x_{n+2} 
+\left(   \prod_{k=1}^{n-1} \sigma_k^z   \right) 
[ \cos (\theta_n) \sin (\theta_{n+1})
\sigma_n^z\sigma^x_{n+1}   \sigma^x_{n+2} 
 - \sin (\theta_n)  \cos (\theta_{n+1})
\sigma^y_n   \sigma^y_{n+1}]\right]
\nonumber 
\end{eqnarray}
or to keep also the interactions between the next-nearest-neighbor LIOMS $\epsilon^z_n $ and $\epsilon^z_{n+2} $ 
\begin{eqnarray}
&& H^{(2)nnn} \equiv \sum_{n=1}^{N-2} J_{n,n+2}^{(2)} \epsilon^z_n \epsilon^z_{n+2}
\label{h2nnn}
\end{eqnarray}
and the interactions between three consecutive LIOMS
\begin{eqnarray}
&& H^{(3)nn} \equiv \sum_{n=1}^{N-2} J_{n,n+1,n+2}^{(3)} \epsilon^z_n \epsilon^z_{n+1} \epsilon^z_{n+2}
\label{h3nn}
\end{eqnarray}

\subsection{ Choice to produce an exact pairing between the two parity sectors $P=\pm 1 $  }

While Eq \ref{hlioms} is the general form of the parent Hamiltonians in terms of the LIOMs $\epsilon_{n=1,..,N-1}^z$ and $\epsilon_N^z=P^z$,
one can also choose to produce an exact pairing in the spectrum between the two parity sectors $P=\pm 1 $
by suppressing all terms containing $P^z$ in Eq. \ref{hlioms}
\begin{eqnarray}
H^{Pairing}
&& =\sum_{p=1}^{N-1}
 \sum_{1 \leq n_1 < n_2 <...< n_p  \leq N-1}  J_{n_1,n_2,..,n_p}^{(p)} \epsilon^z_{n_1}\epsilon^z_{n_2} ... \epsilon^z_{n_p} 
\label{hliomspairing}
\end{eqnarray}
The corresponding $2^{N-1}$ energy-levels parametrized by the binary variables $\epsilon_1=\pm 1,...,\epsilon_{N-1}=\pm 1$
are then all doubly degenerate, since the two MPS $\ket{\psi_{\epsilon_1,...,\epsilon_{N-1} ,P =\pm 1} }$ 
that are related by the operator $\sigma_N^x=P^x$ (Eqs \ref{pairingketp} and \ref{pxparityflip}) have the same energy.
The operator $\sigma_N^x=P^x$ then commutes with the Hamiltonian and anticommute with the parity $P^z$ :
it is thus an exact odd normalized zero-mode.
This notion of odd normalized zero-modes has attracted a lot of interest recently under the name of Majorana Zero Modes (MZM)
in the context of the classification of topological phases \cite{kitaevchain,kitaevfid,kitaevalpha,10phases,topo}.
They have been considered both in random systems in relation with Many-Body-Localization models \cite{maj_pol}
or in non-random models like the integrable XYZ chain \cite{strongzeromode}
where they were called 'Strong Zero Mode',
 with various consequences for the long coherence time of edge spins 
\cite{longcoherence,mila}, for the phenomenon of prethermalization \cite{prethermal},
and for their fate in the presence of dissipation \cite{carollo1},
while generalization to ladders can be found in \cite{carollo2}.
In the Majorana formulation, these exact odd zero modes appear whenever the 
Hamiltonian involves an odd number $(2N-1)$ of Majorana operators 
\cite{akhmerov,goldstein,wilczek,feldman,moreabout,kauffman,c_evenodd}
instead of the even number $(2N)$ of Majorana operators that are needed to describe a chain of $N$ spins (See the reminder in Appendix \ref{app_majorana}).


\section{ Diagonal Ensemble describing the long-time-averaged observables  }

\label{sec_dynamics}

\subsection{ Reminder on the Diagonal Ensemble as a function of the initial density matrix $\rho(t=0)$   }

The unitary dynamics for the density matrix $\rho(t)$
starting from some initial density matrix $\rho(t=0)$  reads
\begin{eqnarray}
\rho(t) = e^{-i H t} \rho(0) e^{i H t} 
\label{dynrho}
\end{eqnarray}
In the presence of continuous disorder, the spectrum is non-degenerate,
and the time-average over a large time-window $[0,t]$
gives the so-called Diagonal Ensemble Density Matrix involving the projectors on the $2^N$ eigenstates
\begin{eqnarray}
\rho^{DE} \equiv \lim_{t \to +\infty} \frac{1}{t} \int_0^t d\tau \rho(\tau) = 
\sum_{\epsilon_1=\pm } ...\sum_{\epsilon_{N-1}=\pm } \sum_{P=\pm }
p^{DE}_{ \epsilon_1,...,\epsilon_{N-1},P } 
\ket{\psi_{\epsilon_1,...,\epsilon_{N-1},P } } 
 \bra{\psi_{\epsilon_1,...,\epsilon_{N-1},P } } 
\label{diagensemble}
\end{eqnarray}
where the weights of the eigenstates have to be computed as a function
of the initial condition $\rho(0)$
\begin{eqnarray}
p^{DE}_{ \epsilon_1,...,\epsilon_{N-1},P } \equiv  \braket { \psi_{\epsilon_1,...,\epsilon_{N-1},P }  \vert \rho(0) \vert  \psi_{\epsilon_1,...,\epsilon_{N-1},P } }
= {\rm Tr} \left(  \rho(0) \ket{\psi_{\epsilon_1,...,\epsilon_{N-1},P } } 
 \bra{\psi_{\epsilon_1,...,\epsilon_{N-1},P } }   \right)
\label{pdiagensemble}
\end{eqnarray}


\subsection{ Analysis of the local magnetizations in the Diagonal Ensemble}

In order to analyze the local magnetization on site $n$  in the Diagonal Ensemble
\begin{eqnarray}
m_n^{DE} \equiv {\rm Tr} \left( \sigma_n^z \rho^{DE} \right)
 = \sum_{\epsilon_1=\pm } ...\sum_{\epsilon_{N-1}=\pm } \sum_{P=\pm }
p^{DE}_{ \epsilon_1,...,\epsilon_{N-1},P }
{\rm Tr} \left( \sigma_n^z 
\ket{\psi_{\epsilon_1,...,\epsilon_{N-1},P } } 
 \bra{\psi_{\epsilon_1,...,\epsilon_{N-1},P } } \right)
\label{mndiagensemble}
\end{eqnarray}
it is simpler to work in the basis of pseudo-spin operators,
where the projectors read
\begin{eqnarray}
\ket{\psi_{\epsilon_1,...,\epsilon_{N-1},P } } 
 \bra{\psi_{\epsilon_1,...,\epsilon_{N-1},P } } = 
\left( \frac{1+ P P^z}{2}  \right) \prod_{k=1}^{N-1}\left( \frac{1+ \epsilon_k \epsilon_k^z}{2}  \right)
\label{diagensembleepseudo}
\end{eqnarray}
while $\sigma_n^z$ has already been computed in Eq. \ref{inversionz}
for $1<n<N$
\begin{eqnarray}
\sigma_n^z   && =  
\left[ \cos (\theta_{n-1}) \epsilon_{n-1}^z+\sin (\theta_{n-1}) \epsilon_{n-1}^x \right]
\left[ \cos (\theta_n) \epsilon_n^z+\sin (\theta_n) \epsilon_n^x \right]
\label{inversionzbis}
\end{eqnarray}
so that one obtains the very simple result
\begin{eqnarray}
{\rm Tr} \left( \sigma_n^z 
\ket{\psi_{\epsilon_1,...,\epsilon_{N-1},P } } 
 \bra{\psi_{\epsilon_1,...,\epsilon_{N-1},P } } \right)
=\epsilon_{n-1} \epsilon_n \cos (\theta_{n-1})\cos (\theta_n)
\label{tracemneps}
\end{eqnarray}
Then the magnetization of Eq. \ref{mndiagensemble} becomes using the weights of Eq. \ref{pdiagensemble} and the projectors of Eq. \ref{diagensembleepseudo}
\begin{eqnarray}
m_n^{DE} 
 && = \cos (\theta_{n-1})\cos (\theta_n)
 \sum_{\epsilon_1=\pm } ...\sum_{\epsilon_{N-1}=\pm } \sum_{P=\pm }
\epsilon_{n-1} \epsilon_n
p^{DE}_{ \epsilon_1,...,\epsilon_{N-1},P }
\nonumber \\
&&= \cos (\theta_{n-1})\cos (\theta_n)
 {\rm Tr} \left(  \rho(0) 
\left[ \sum_{\epsilon_1=\pm } ...\sum_{\epsilon_{N-1}=\pm } \sum_{P=\pm }
\epsilon_{n-1} \epsilon_n\ket{\psi_{\epsilon_1,...,\epsilon_{N-1},P } } 
 \bra{\psi_{\epsilon_1,...,\epsilon_{N-1},P } }  \right] \right)
\nonumber \\
&&
=
 \cos (\theta_{n-1})\cos (\theta_n)
 {\rm Tr} \left(  \rho(0) \epsilon_{n-1}^z \epsilon_n^z   \right)
\label{mndiagensembleres}
\end{eqnarray}
The translation of the pseudo-spin operators $\epsilon_{k}^z $ (Eq. \ref{enzsigma})
in terms of the initial spin operators $\sigma_m^{x,y,z}$ yields 
\begin{eqnarray}
\epsilon_{n-1}^z \epsilon_n^z 
&&=  \cos (\theta_{n-1})  \cos (\theta_{n}) \sigma^z_{n} 
+\sin (\theta_{n-1}) \sin (\theta_{n})\sigma^x_{n-1}    \sigma^x_{n+1} 
\nonumber \\
&&+\cos (\theta_{n-1}) \sin (\theta_{n})
\left(   \prod_{k=1}^{n-1} \sigma_k^z   \right) \sigma^x_{n}   \sigma^x_{n+1} 
 - \sin (\theta_{n-1})  \cos (\theta_{n})
\left(   \prod_{k=1}^{n-2} \sigma_k^z   \right)
\sigma^y_{n-1}   \sigma^y_{n}  
\label{enzenznext}
\end{eqnarray}
that determines which operators in the initial density matrix $\rho(0)$
are involved in the magnetization $m_n^{DE}$ of Eq. \ref{mndiagensemble}.
To be more concrete, let us consider two simple examples for the initial condition at $t=0$ :

(i) if the initial condition is fully magnetized along the direction $z$ with magnetization $S_n$ on site $n$
\begin{eqnarray}
\ket {\psi(0)}=\ket {\sigma^z_1=S_1 }\ket {\sigma^z_2=S_2 } ... \ket {\sigma^z_N=S_N }
\label{ketinisigmaz}
\end{eqnarray}
the initial density matrix 
\begin{eqnarray}
\rho(0) =  \prod_{k=1}^{N}\left( \frac{1+ S_k \sigma_k^z}{2}  \right)
\label{rhoinisigmaz}
\end{eqnarray}
yields that the magnetization of the Diagonal Ensemble of Eq. \ref{mndiagensembleres}
\begin{eqnarray}
m_n^{DE} = \cos^2 (\theta_{n-1})\cos^2 (\theta_n) S_n
\label{mndiagensembleresiniz}
\end{eqnarray}
keeps the memory of the initial magnetization $S_n$ even if it is reduced in amplitude
by the angles $(\theta_{n-1},\theta_{n})$ of the two neighboring bonds.

(ii) if the initial condition is instead fully magnetized along the direction  $x$ with magnetization $\zeta_n$ on site $n$
\begin{eqnarray}
\ket {\psi(0)}=\ket {\sigma^x_1=\zeta_1 }\ket {\sigma^x_2=\zeta_2 } ... \ket {\sigma^x_N=\zeta_N }
\label{ketinisigmax}
\end{eqnarray}
the initial density matrix 
\begin{eqnarray}
\rho(0) =  \prod_{k=1}^{N}\left( \frac{1+ \zeta_k \sigma_k^x}{2}  \right)
\label{rhoinisigmax}
\end{eqnarray}
yields that the magnetization of the Diagonal Ensemble of Eq. \ref{mndiagensembleres}
\begin{eqnarray}
m_n^{DE} = \cos (\theta_{n-1})\cos (\theta_n) \sin (\theta_{n-1}) \sin (\theta_{n})
\zeta_{n-1} \zeta_{n+1}
\label{mndiagensembleresinix}
\end{eqnarray}
is non-vanishing and keeps the memory of the $\sigma^x$-magnetizations
on the two neighboring sites $ \zeta_{n\pm 1}$.

More generally, one can compute along the same lines
the values of local operators in the Diagonal Ensemble as a function of the initial condition $\rho(t=0)$.



\subsection{Operator-Space-Entanglement of the Diagonal Ensemble for the initial condition $\ket {\psi(0)}=\ket{ S_1,...,S_N}  $}

When the initial condition is given by Eq. \ref{ketinisigmaz},
 the weights of the eigenstates in the Diagonal Ensemble
of Eq .\ref{pdiagensemble} have already been evaluated in Eq. \ref{mpscoef}
\begin{eqnarray}
p^{DE}_{ \epsilon_1,...,\epsilon_{N-1},P } =
\vert  \braket{ S_1,...,S_N \vert \psi_{\epsilon_1,...,\epsilon_{N-1},P  }} \vert^2
= \delta_{P, \prod_{j=1}^N S_j} \prod_{k=1}^{N-1} 
\left( \lambda^{[\epsilon_k] \prod_{n=1}^k S_n }_k \right)^2
\label{pdiagensembleinispins}
\end{eqnarray}
The statistics of these $2^N$ weights $p^{DE}_{ \epsilon_1=\pm 1,...,\epsilon_{N-1}=\pm 1,P=\pm }  $ normalized to unity 
can be analyzed via the Inverse Participation Ratios as a function of the continuous parameter $q$
\begin{eqnarray}
{\cal Y}^{(q)}_{S_1,..,S_N} && \equiv \sum_{\epsilon_1 =\pm} ...  \sum_{\epsilon_{N-1} =\pm} \sum_{P =\pm}
\left( p^{DE}_{ \epsilon_1,...,\epsilon_{N-1},P } \right)^q 
=  \prod_{k=1}^{N-1} 
\left[ \sum_{\epsilon_k =\pm} 
\left( \lambda^{[\epsilon_k] \prod_{n=1}^k S_n }_k \right)^{2q} \right]
 \nonumber \\ &&
 = \prod_{k=1}^{N-1} \left(  \cos^{2q} \left( \frac{\theta_{k}}{2} \right)
+\sin^{2q} \left( \frac{\theta_{k}}{2} \right)
\right)
\label{yqdiag}
\end{eqnarray}
They are thus independent on the precise values 
$S_n=\pm 1$ of the spins in initial state $\ket {\psi(0)}=\ket{ S_1,...,S_N}  $
and coincide with the values of Eq. \ref{yq}.
In particular, the value for $q=2$ represents the purity ${\cal P}^{DE}$ of the Diagonal Ensemble
density matrix $\rho^{DE} $
\begin{eqnarray}
{\cal P}^{DE} \equiv {\rm Tr}_{[1,..,N]} \left[ (\rho^{DE})^2 \right] && =\sum_{\epsilon_1 =\pm} ...  \sum_{\epsilon_{N-1} =\pm} \sum_{P =\pm}
\left( p^{DE}_{ \epsilon_1,...,\epsilon_{N-1},P } \right)^2
=
{\cal Y}^{(q=2)}_{S_1,..,S_N} 
 = \prod_{k=1}^{N-1} \left[  \cos^{4} \left( \frac{\theta_{k}}{2} \right)
+\sin^{4} \left( \frac{\theta_{k}}{2} \right)
\right]
\nonumber \\
&& =\prod_{k=1}^{N-1} \left( \frac{1+ \cos ^2 \theta_k}{2} \right)
\label{puritydiag}
\end{eqnarray}

The product form of the weights in Eq. \ref{pdiagensembleinispins}
and the MPO forms of the MPS-projectors (Eq. \ref{mpoproj})
\begin{eqnarray}
\ket{\psi_{\epsilon_1,...,\epsilon_{N-1},P } } 
 \bra{\psi_{\epsilon_1,...,\epsilon_{N-1},P } } 
= \sum_{ \substack{\alpha_1 =\pm \\ \alpha_1'=\pm} }
...
\sum_{ \substack{\alpha_{N-1} =\pm \\ \alpha_{N-1}'=\pm} }
\left[ \prod_{k=1}^{N-1}  
\lambda^{[\epsilon_k] \alpha_k}_k 
\lambda^{[\epsilon_k] \alpha_k'}_k  
 \right]
O_1^{  \alpha_1, \alpha_1'} 
\left[ \prod_{k=2}^{N-1} O_k^{ \alpha_{k-1} \alpha_k,  \alpha_{k-1}' \alpha_k'} 
\right]
O_N^{ \alpha_{N-1} P , \alpha_{N-1}' P }
\label{mpoprojdiag}
\end{eqnarray}
yields the following MPO form for the Diagonal Ensemble density matrix of Eq. \ref{diagensemble}
\begin{eqnarray}
\rho^{DE}  && = 
 \sum_{ \substack{\alpha_1 =\pm \\ \alpha_1'=\pm} }
...
\sum_{ \substack{\alpha_{N-1} =\pm \\ \alpha_{N-1}'=\pm} }
\left[ \prod_{k=1}^{N-1} 
G_k^{\alpha_k,\alpha_k'}
 \right]
O_1^{  \alpha_1, \alpha_1'} 
\left[ \prod_{k=2}^{N-1} O_k^{ \alpha_{k-1} \alpha_k,  \alpha_{k-1}' \alpha_k'} 
\right]
 O_N^{ \alpha_{N-1}  \prod_{j=1}^N S_j, \alpha_{N-1}' \prod_{j=1}^N S_j }
\label{diagensemblempo}
\end{eqnarray}
with the bond variables
\begin{eqnarray}
G_k^{\alpha_k,\alpha_k'} && \equiv 
\sum_{\epsilon_k=\pm }\left( \lambda^{[\epsilon_k] \prod_{n=1}^k S_n }_k \right)^2
\lambda^{[\epsilon_k] \alpha_k}_k 
\lambda^{[\epsilon_k] \alpha_k'}_k  
\nonumber \\
&& = \delta_{\alpha_k,\alpha_k'} \frac{1+ \alpha_k (\prod_{n=1}^k S_n ) \cos^2 \theta_k }{2}
+ \delta_{\alpha_k,-\alpha_k'} \frac{ (\prod_{n=1}^k S_n ) \cos \theta_k \sin \theta_k}{2}
\label{gkaa}
\end{eqnarray}

In particular, the decomposition across the bond $(n,n+1)$
\begin{eqnarray}
\rho^{DE}  && = \sum_{ \substack{\alpha_n =\pm \\ \alpha_n'=\pm} }
 {\cal L}^{\alpha_n,\alpha_n'}_{[1,..,n]}  \ G_n^{\alpha_n,\alpha_n'} \ {\cal R}^{\alpha_n,\alpha_n'}_{[n+1,..,N]}
\label{rhodiagLR}
\end{eqnarray}
involves the Left operator that resums the MPO for the Left part $[1,..,n]$
\begin{eqnarray}
 {\cal L}^{\alpha_n,\alpha_n'}_{[1,..,n]}   && = 
 \sum_{ \substack{\alpha_1 =\pm \\ \alpha_1'=\pm} }
...
\sum_{ \substack{\alpha_{n-1} =\pm \\ \alpha_{n-1}'=\pm} }
\left[ \prod_{k=1}^{n-1} 
G_k^{\alpha_k,\alpha_k'}
 \right]
O_1^{  \alpha_1, \alpha_1'} 
\left[ \prod_{k=2}^{n} O_k^{ \alpha_{k-1} \alpha_k,  \alpha_{k-1}' \alpha_k'} 
\right]
\label{rhodiagL}
\end{eqnarray}
and the Right operator that resums the MPO for the Right part $[n+1,..,N]$
\begin{eqnarray}
 {\cal R}^{\alpha_n,\alpha_n'}_{[n+1,..,N]} && = 
 \sum_{ \substack{\alpha_{n+1} =\pm \\ \alpha_{n+1}'=\pm} }
...
\sum_{ \substack{\alpha_{N-1} =\pm \\ \alpha_{N-1}'=\pm} }
\left[ \prod_{k=n+1}^{N-1} 
G_k^{\alpha_k,\alpha_k'}
 \right]
\left[ \prod_{k=n+1}^{N-1} O_k^{ \alpha_{k-1} \alpha_k,  \alpha_{k-1}' \alpha_k'} 
\right]
 O_N^{ \alpha_{N-1}  \prod_{j=1}^N S_j, \alpha_{N-1}' \prod_{j=1}^N S_j }
\label{rhodiagR}
\end{eqnarray}

With respect to the Hilbert-Schmidt inner-product for operators on the Left part $[1,..,n]$
and for operators on the Right part $[n+1,..,N]$, these operators satisfy the orthogonality properties
\begin{eqnarray}
\left({\cal L}^{\alpha_n,\alpha_n'}_{[1,..,n]}  \vert {\cal L}^{\beta_n,\beta_n'}_{[1,..,n]} \right)_{HS[1,..,n]}  
&& \equiv {\rm Tr}_{[1,..,n]} \left[ \left( {\cal L}^{\alpha_n,\alpha_n'}_{[1,..,n]} \right)^{\dagger}
 {\cal L}^{\beta_n,\beta_n'}_{[1,..,n]} \right]   = \delta_{\alpha_n, \beta_n} \delta_{\alpha_n', \beta_n'} 
\left\vert \left\vert {\cal L}^{\alpha_n,\alpha_n'}_{[1,..,n]} \right\vert \right\vert^2_{HS[1,..,n]} 
\label{rhodiagLortho} \\
\left({\cal R}^{\alpha_n,\alpha_n'}_{[n+1,..,N]}  \vert {\cal R}^{\beta_n,\beta_n'}_{[n+1,..,N]} \right)_{HS[n+1,..,N]}  
&& \equiv {\rm Tr}_{[n+1,..,N]} \left[ \left( {\cal R}^{\alpha_n,\alpha_n'}_{[n+1,..,N]} \right)^{\dagger}
 {\cal R}^{\beta_n,\beta_n'}_{[n+1,..,N]} \right]   = \delta_{\alpha_n, \beta_n} \delta_{\alpha_n', \beta_n'} 
\left\vert \left\vert {\cal R}^{\alpha_n,\alpha_n'}_{[n+1,..,N]} \right\vert \right\vert^2_{HS[n+1,..,N]} 
\nonumber
\end{eqnarray}
while their Hilbert-Schmidt squared norms can be computed using the explicit expression of the bond variables (Eq. \ref{gkaa})
\begin{eqnarray}
\left\vert \left\vert {\cal L}^{\alpha_n,\alpha_n'}_{[1,..,n]} \right\vert \right\vert^2_{HS[1,..,n]} 
 && \equiv 
{\rm Tr}_{[1,..,n]} \left[ \left( {\cal L}^{\alpha_n,\alpha_n'}_{[1,..,n]} \right)^{\dagger}
 {\cal L}^{\alpha_n,\alpha_n'}_{[1,..,n]} \right] 
=\prod_{k=1}^{n-1}  \left[ \sum_{ \substack{\alpha_k =\pm \\ \alpha_k'=\pm} }
\left( G_k^{\alpha_k,\alpha_k'} \right)^2
 \right]
\nonumber \\
&& =\prod_{k=1}^{n-1}  \left[ \frac{1+ \cos^2 \theta_k}{2}  \right]
\equiv {\cal P}_{(1,..,n-1)}
\nonumber \\
\left\vert \left\vert {\cal R}^{\alpha_n,\alpha_n'}_{[n+1,..,N]} \right\vert \right\vert^2_{HS[n+1,..,N]} 
 && \equiv 
{\rm Tr}_{[n+1,..,N]} \left[ \left( {\cal R}^{\alpha_n,\alpha_n'}_{[n+1,..,N]} \right)^{\dagger}
 {\cal R}^{\alpha_n,\alpha_n'}_{[n+1,..,N]} \right] 
=\prod_{k=n+1}^{N-1}  \left[ \sum_{ \substack{\alpha_k =\pm \\ \alpha_k'=\pm} }
\left( G_k^{\alpha_k,\alpha_k'} \right)^2
 \right]
\nonumber \\
&&=\prod_{k=n+1}^{N-1}  \left[ \frac{1+ \cos^2 \theta_k}{2} \right]\equiv {\cal P}_{(n+1,..,N-1)}
\label{hsnorm}
\end{eqnarray}
These norms do not depend on the indices $(\alpha_n=\pm1,\alpha_n'=\pm 1)$
and correspond to the contribution ${\cal P}_{(1,..,n-1)} $ from the Left part $[1,..,n]$ 
and to the contribution ${\cal P}_{(n+1,..,N-1)} $ from the Right part $[n+1,..,N]$
in the total purity $  {\cal P}^{DE} $ of Eq. \ref{puritydiag} for the Diagonal Ensemble
that can be rewritten as
\begin{eqnarray}
{\cal P}^{DE}  =\prod_{k=1}^{N-1} \left( \frac{1+ \cos ^2 \theta_k}{2} \right) = 
{\cal P}_{(1,..,n-1)}\left( \frac{1+ \cos ^2 \theta_n}{2} \right){\cal P}_{(n+1,..,N-1)}
\label{puritydiagLR}
\end{eqnarray}

The notion of Operator-Space-Entanglement for density matrices describing mixed states \cite{mpdo,mpd,prosen,zhou,whiteR,zilber}
can be applied to the Diagonal Ensemble density matrix $\rho^{DE} $ of Eq. \ref{rhodiagLR}
by considering the operator
\begin{eqnarray}
\left( \rho^{DE} \right)^{\dagger} \rho^{DE}  && =
 \sum_{ \substack{\alpha_n =\pm \\ \alpha_n'=\pm} }
 \sum_{ \substack{\beta_n =\pm \\ \beta_n'=\pm} }
\left( {\cal L}^{\alpha_n,\alpha_n'}_{[1,..,n]} \right)^{\dagger}{\cal L}^{\beta_n,\beta_n'}_{[1,..,n]} 
 \ G_n^{\alpha_n,\alpha_n'} G_n^{\beta_n,\beta_n'} 
\ \left( {\cal R}^{\alpha_n,\alpha_n'}_{[n+1,..,N]} \right)^{\dagger} {\cal R}^{\beta_n,\beta_n'}_{[n+1,..,N]}
\label{rhodiagLRdagger}
\end{eqnarray}
Its trace over the full chain $[1,..,N]$ corresponds to
the Hilbert-Schmidt norm over the full chain $[1,..,N]$ and to the purity of Eq. \ref{puritydiag} since 
the density matrix is hermitian $\left(\rho^{DE} \right)^{\dagger}=\rho^{DE}  $
\begin{eqnarray}
\vert \vert \rho^{DE} \vert \vert^2_{HS[1,..,n]} && \equiv 
{\rm Tr}_{[1,..,n]} \left[ 
\left( \rho^{DE} \right)^{\dagger} \rho^{DE} 
 \right] 
={\rm Tr}_{[1,..,n]} \left[ 
\left( \rho^{DE} \right)^{2} 
 \right] ={\cal P}^{DE} 
\label{purityHS}
\end{eqnarray}
When one evaluates only the partial trace over the Left Part $(n+1,..,N)$
or only
the partial trace over the Right Part $(n+1,..,N)$, one obtains 
using Eqs \ref{rhodiagLortho} and \ref{hsnorm}
that they are diagonal in their respective operator spaces
\begin{eqnarray}
{\rm Tr}_{[1,..,n]} \left[ \left( \rho^{DE} \right)^{\dagger} \rho^{DE} \right] && =
 \sum_{ \substack{\alpha_n =\pm \\ \alpha_n'=\pm} }
 \sum_{ \substack{\beta_n =\pm \\ \beta_n'=\pm} }
{\rm Tr}_{[1,..,n]}\left[ \left( {\cal L}^{\alpha_n,\alpha_n'}_{[1,..,n]} \right)^{\dagger}{\cal L}^{\beta_n,\beta_n'}_{[1,..,n]} 
\right]
 \ G_n^{\alpha_n,\alpha_n'} G_n^{\beta_n,\beta_n'} 
\ \left( {\cal R}^{\alpha_n,\alpha_n'}_{[n+1,..,N]} \right)^{\dagger} {\cal R}^{\beta_n,\beta_n'}_{[n+1,..,N]}
\nonumber \\
&& = {\cal P}_{(1,..,n-1)} \sum_{ \substack{\alpha_n =\pm \\ \alpha_n'=\pm} }
 \ \left( G_n^{\alpha_n,\alpha_n'} \right)^2
\ \left( {\cal R}^{\alpha_n,\alpha_n'}_{[n+1,..,N]} \right)^{\dagger} {\cal R}^{\alpha_n,\alpha_n'}_{[n+1,..,N]}
\nonumber \\
{\rm Tr}_{[n+1,..,N]} \left[ \left( \rho^{DE} \right)^{\dagger} \rho^{DE} \right] && =
 \sum_{ \substack{\alpha_n =\pm \\ \alpha_n'=\pm} }
 \sum_{ \substack{\beta_n =\pm \\ \beta_n'=\pm} }
\left( {\cal L}^{\alpha_n,\alpha_n'}_{[1,..,n]} \right)^{\dagger}{\cal L}^{\beta_n,\beta_n'}_{[1,..,n]} 
 \ G_n^{\alpha_n,\alpha_n'} G_n^{\beta_n,\beta_n'} 
\ {\rm Tr}_{[n+1,..,N]} \left[ \left( {\cal R}^{\alpha_n,\alpha_n'}_{[n+1,..,N]} \right)^{\dagger} {\cal R}^{\beta_n,\beta_n'}_{[n+1,..,N]} \right]
\nonumber \\
&& = {\cal P}_{(n+1,..,N-1)} \sum_{ \substack{\alpha_n =\pm \\ \alpha_n'=\pm} }
\left( G_n^{\alpha_n,\alpha_n'} \right)^2
\left( {\cal L}^{\alpha_n,\alpha_n'}_{[1,..,n]} \right)^{\dagger}{\cal L}^{\alpha_n,\alpha_n'}_{[1,..,n]} 
\label{rhodiagLRdaggerpartialtrace}
\end{eqnarray}
where, besides the partial purity factors ${\cal P}_{(1,..,n-1)}  $ and $ {\cal P}_{(n+1,..,N-1)} $
already discussed in Eq. \ref{hsnorm},
 the four common eigenvalues for $\alpha_n =\pm $ and $\alpha_n'=\pm $ 
are given by the squares of the bond variable of Eq. \ref{gkaa}
\begin{eqnarray}
\left( G_n^{\alpha_n,\alpha_n'} \right)^2
 = \delta_{\alpha_n,\alpha_n'} \frac{1+ \cos^4 \theta_n+ 2 \alpha_n (\prod_{k=1}^n S_k ) \cos^2 \theta_n }{4}
+ \delta_{\alpha_n,-\alpha_n'} \frac{  \cos^2 \theta_n \sin^2 \theta_n}{4}
\label{gkaasquare}
\end{eqnarray}
whose sum reproduces the missing factor related to the bond $(n,n+1)$ in the total purity of Eq. \ref{puritydiagLR}
\begin{eqnarray}
\sum_{ \substack{\alpha_n =\pm \\ \alpha_n'=\pm} }\left( G_n^{\alpha_n,\alpha_n'} \right)^2
 = \frac{1+ \cos^2 \theta_n }{2}
\label{gkaasquaresum}
\end{eqnarray}
So the MPO form of Eq. \ref{diagensemblempo}
for the Diagonal Ensemble $\rho^{DE}$
is the analog at the level of operators of the Vidal canonical form for MPS,
where all the entanglement properties for any bipartition in two parts $[1,..,n]$ and $[n+1,..,N]$
are directly accessible.


\section{ Conclusion  }

\label{sec_conclusion}

In this paper, we have considered the inverse problem of  'eigenstates-to-Hamiltonian' in the context of Many-Body-Localization for an open chain of $N$ quantum spins. We have first constructed the simplest orthonormal basis of the Hilbert space made of $2^N$ Matrix-Product-States (MPS) of bond dimension $D=2$, that have all 
the same entanglement entropy across each bond and have all the same multifractal dimensions.
We have then analyzed the corresponding pseudo-spin operators that can be considered as the local building blocks of these $2^N$ MPS, in order to construct the parent Hamiltonians that have these $2^N$ MPS as eigenstates. Finally we 
have studied the Matrix-Product-Operator form of the Diagonal Ensemble density matrix $\rho^{DE}$ that allows to compute long-time-averaged observables of the unitary dynamics. We have given explicit results for the memory of the local magnetizations as a function of the initial density matrix $\rho(t=0)$. Finally, we have studied the entanglement properties of the Diagonal Ensemble density matrix $\rho^{DE}$ in operator-space, via the generalized notion of Schmidt decomposition for density matrices describing mixed states.

Our conclusion is that the explicit construction of Many-Body-Localized models via this 'eigenstates-to-Hamiltonian'
 inverse procedure provides a more concrete picture of the Local Integrals of Motion, of the memory effects
in local observables, and of the entanglement structure in the Diagonal Ensemble. In the future, 
it would be interesting to build similarly other explicit models both in dimension $d=1$ and in higher dimension $d>1$.


\appendix

\section{ Interpretation of the Local Integrals of Motion (LIOMs) in terms of Majorana fermions }

\label{app_majorana}

\subsection{ Reminder on the translation of the initial spin operators $\sigma_k^{x,y,z}$ in terms of Majorana operators $\gamma_j$ }

The $(2N)$ Majorana operators 
\begin{eqnarray}
\gamma_{2j-1} \equiv a_j && \equiv \left( \prod_{k=1}^{j-1} \sigma_k^z \right) \sigma_j^x
\nonumber \\
\gamma_{2j} \equiv b_j && \equiv  \left( \prod_{k=1}^{j-1} \sigma_k^z \right)  \sigma_j^y 
\label{sigmaxy}
\end{eqnarray}
 are hermitian
\begin{eqnarray}
\gamma_j^{\dagger}=\gamma_j
\label{hermi}
\end{eqnarray}
 square to the Identity
\begin{eqnarray}
\gamma_j^2=  \mathbb{1}
\label{squareunity}
\end{eqnarray}
and anti-commute with each other
\begin{eqnarray}
\{ \gamma_j , \gamma_l \} \equiv \gamma_j \gamma_l + \gamma_l \gamma_j && = 0 \ \ \ \ \ \  { \rm for } \ \ \   j \ne l
\label{anticomm}
\end{eqnarray}

Depending on the circumstances, one may prefer the unifying writing in terms of the $(2N)$ Majorana fermions $\gamma_{j=1,..,2N}$
(as in the anticommuting relations of Eq. \ref{anticomm}),
or one may prefer to use the notation with two flavors $a_{j=1,..,N}$ and $b_{j=1,..,N}$ in order stress their different behaviors
with respect to the Time-Reversal-Symmetry $T$
\begin{eqnarray}
T i T^{-1} && =  -i
\nonumber \\
T a_j T^{-1} &&=  a_j
\nonumber \\
T b_j T^{-1} && = - b_j
\label{time}
\end{eqnarray}
Reciprocally, the operator $\sigma_j^z$ corresponds to the pairing of two Majorana operators on the same site $j$
\begin{eqnarray}
 \sigma^z_j && =  -i  \gamma_{2j-1} \gamma_{2j} = -i a_j b_j
\label{dicoz}
\end{eqnarray}
while the operators $\sigma_j^{x,y} $ correspond to the strings of operators
\begin{eqnarray}
\sigma_j^x && = \left( \prod_{k=1}^{j-1} ( -i  \gamma_{2k-1} \gamma_{2k})  \right) \gamma_{2j-1} 
= \left( \prod_{k=1}^{j-1} ( -i  a_k b_k)  \right) a_j
\nonumber \\
\sigma_j^y && = \left( \prod_{k=1}^{j-1} ( -i  \gamma_{2k-1} \gamma_{2k}) \right)\gamma_{2j} 
= \left( \prod_{k=1}^{j-1} ( -i  a_k b_k)  \right) b_j
\label{sigmaxymaj}
\end{eqnarray}
The two-spin operators $xx$ and $yy$ corresponds to the pairing of two Majorana operators belonging to two neighboring sites $j$ and $j+1$
\begin{eqnarray}
\sigma_j^x \sigma^x_{j+1} && = -i    \gamma_{2j} \gamma_{2j+1} =-i b_j a_{j+1}
\nonumber \\
\sigma_j^y \sigma^y_{j+1}  && = i  \gamma_{2j-1}  \gamma_{2j+2} = i a_j b_{j+1}
\label{dico}
\end{eqnarray}
while the two-spin operator $zz$ corresponds to the interaction between the four Majorana operators of two consecutive sites $j$ and $j+1$
\begin{eqnarray}
\sigma_j^z \sigma^z_{j+1} && = -\gamma_{2j-1} \gamma_{2j} \gamma_{2j+1} \gamma_{2j+2} = - a_j b_j a_{j+1} b_{j+1}
\label{dicozz}
\end{eqnarray}


\subsection{ Interpretation of the LIOMs $\epsilon_n^{z}$ in terms of pseudo-Majorana operators  }

For $1 \leq n \leq N-1$, the pseudo-spin operator $\epsilon_n^{z}$ of Eq \ref{enzsigma} 
can be translated in terms of the Majorana operators introduced above 
\begin{eqnarray}
\epsilon_n^z
&&=  \cos (\theta_n) 
\left(   \prod_{k=1}^{n-1} (-i  \gamma_{2k-1} \gamma_{2k} )   \right)
(-i  \gamma_{2n-1} \gamma_{2n} ) 
-i \sin (\theta_n)     \gamma_{2n} \gamma_{2n+1}
\nonumber \\
&& =
-i \left[  \cos (\theta_n) 
\left(   \prod_{k=1}^{n-1} ( -i a_k b_k)   \right)  a_n 
- \sin (\theta_n)   a_{n+1} \right] b_n
\equiv -i {\tilde a}_n b_n
\label{liommajorana}
\end{eqnarray}
and can be thus interpreted as the pairing between the Majorana operator $b_n$
with the new pseudo-Majorana operator
\begin{eqnarray}
{\tilde a}_n
&& \equiv   \cos (\theta_n) \left(   \prod_{k=1}^{n-1} ( -i a_k b_k)   \right)  a_n  - \sin (\theta_n)   a_{n+1} 
\label{newamajorana}
\end{eqnarray}
i.e more explicitly for $n=1,2,3..,N-1$
\begin{eqnarray}
{\tilde a}_1  &&=   \cos (\theta_1)   a_1 - \sin (\theta_1)   a_2
\nonumber \\
{\tilde a}_2  &&=   \cos (\theta_2)  ( -i a_1 b_1)    a_2 - \sin (\theta_2)   a_3 
\nonumber \\
{\tilde a}_3 &&=   \cos (\theta_3) (  - a_1 b_1 a_2 b_2)   a_3 - \sin (\theta_3)   a_{4} 
\nonumber \\
... && ...
\nonumber \\
{\tilde a}_{N-1}
&&=   \cos (\theta_{N-1}) \left(   \prod_{k=1}^{N-2} ( -i a_k b_k)   \right)  a_{N-1} 
- \sin (\theta_{N-1})   a_N
\label{newamajoranaexpli}
\end{eqnarray}

Similarly, the parity operator of Eq. \ref{paritysigmaz}
\begin{eqnarray}
P^z   =\prod_{k=1}^N ( -i  a_k b_k ) \equiv -i {\tilde a}_N b_N
\label{paritytotaltilde}
\end{eqnarray}
can be interpreted as the pairing between the Majorana operator $b_N$
and the new pseudo-Majorana operator
\begin{eqnarray}
{\tilde a}_N = \left( \prod_{k=1}^{N-1} ( -i  a_k b_k ) \right) a_N = \sigma_N^x
\label{newmajoranalast}
\end{eqnarray}

In conclusion, the Local Integrals of Motion (LIOMs),
i.e. the pseudo-spins operators $\epsilon_{k=1,..,N-1}^z$ and the Parity $P^z$,
can be interpreted in terms of the pairing between the $N$ initial Majorana fermions $b_{n=1,..,N}$
with the new pseudo-Majorana fermions ${\tilde a}_n $ defined above.



\begin{thebibliography}{99}






 \bibitem{nandkishore15}
R. Nandkishore and D. A. Huse, 
Ann. Review of Cond. Mat. Phys. 6, 15 (2015).

 \bibitem{altman15mblreview}
 E. Altman and R. Vosk, 
Ann. Review of Cond. Mat. Phys. 6, 383 (2015).

\bibitem{parameswaran17}
S. A. Parameswaran, A. C. Potter and R. Vasseur, 
Annalen der Physik  529, 1600302 (2017).

\bibitem{review_mblergo}
D. J. Luitz, Y. Bar Lev, Annalen der Physik  1600350 (2017)

\bibitem{review_prelovsek}
P. Prelovsek, M. Mierzejewski, O. Barisic, J. Herbrych, Annalen der Physik  1600362 (2017)

\bibitem{review_rare}
K. Agarwal {\it et al}, Annalen der Physik  1600326 (2017)


 \bibitem{alet18}
F. Alet and N. Laflorencie, 
Comptes Rendus Physique (2018).

\bibitem{abanin18}
D. A. Abanin, E. Altman, I. Bloch and M. Serbyn,
Rev. Mod. Phys. 91, 021001 (2019)


\bibitem{bauer}
B. Bauer and C. Nayak, J. Stat. Mech. P09005 (2013).

\bibitem{kjall}
J. A. Kj\"all, J. H. Bardarson and F. Pollmann, Phys. Rev. Lett. 113, 107204 (2014).

\bibitem{alet}
D. J. Luitz, N. Laflorencie and F. Alet, Phys. Rev. B 91, 081103 (2015).

\bibitem{lim1}
S. P. Lim and D. N. Sheng
Phys. Rev. B 94, 045111 (2016)

\bibitem{lim2}
V. Khemani, S. P. Lim, D. N. Sheng, and D. A. Huse
Phys. Rev. X 7, 021013 (2017).


 
\bibitem{pekker1}
D. Pekker and B.K. Clark, Phys. Rev. B 95, 035116 (2017).



\bibitem{pekker2}
X. Yu, D. Pekker and B.K. Clark, Phys. Rev. Lett. 118, 017201 (2017)


\bibitem{friesdorf}
M. Friesdorf, A.H. Werner, W. Brown, V. B. Scholz and J. Eisert, Phys. Rev. Lett. 114, 170505 (2015).

\bibitem{tensor}
A. Chandran, J. Carrasquilla, I.H. Kim, D.A. Abanin and G. Vidal, Phys. Rev. B 92, 024201 (2015).

\bibitem{sondhi}
V. Khemani, F. Pollmann and S. L. Sondhi, Phys. Rev. Lett. 116, 247204 (2016).

\bibitem{dmrgx}
T. Devakul, V. Khemani, F. Pollmann, D. Huse and S. Sondhi, Phil. Trans. of the Roy. Soc. of London A, 375, 2108 (2017)

\bibitem{dmrgxfloquet}
 C. Zhang, F. Pollmann, S. L. Sondhi, R. Moessner, Annalen der Physik 529.7 (2017)


\bibitem{white1} 
S. R. White, Phys. Rev. Lett. 69, 2863 (1992)

\bibitem{white2}
S. R. White, Phys. Rev. B 48, 10345 (1992).

\bibitem{ulrich2005}
U. Schollwoeck, Rev. Mod. Phys. 77, 259 (2005).


\bibitem{pekker14rsrgx}
D. Pekker, G. Refael, E. Altman, E. Demler and V. Oganesyan, 
Phys. Rev. X 4, 011052 (2014).

 
 
\bibitem{huang14} 
Y. Huang, Joel E. Moore, 
Phys. Rev B90, 220202 (2014).



 \bibitem{pouranvari15}
M. Pouranvari and K. Yang, 
Phys. Rev. B 92, 245134 (2015).


\bibitem{agarwal15} 
 K. Agarwal, E. Demler and I. Martin, 
Phys. Rev. B 92, 184203 (2015).



 \bibitem{vasseur15hot}
R. Vasseur, A. C. Potter and S.A. Parameswaran, 
Phys. Rev. Lett. 114, 217201 (2015).


\bibitem{monthus16emergent}
C. Monthus,
  J. Stat. Mech.  033101 (2016).


\bibitem{vasseur16particlehole}
    R. Vasseur, A. J. Friedman, S. A. Parameswaran and A. C. Potter, 
  Phys. Rev. B 93, 134207 (2016).


\bibitem{slage16}
K. Slagle, Y. Z. You, and C. Xu, 
Phys. Rev. B 94, 014205 (2016).




\bibitem{you16}
 Y.Z. You, X.L. Qi and C. Xu, 
Phys. Rev. B 93, 104205 (2016).
 

\bibitem{kang17} 
B. Kang, A. C. Potter and R. Vasseur,
Phys. Rev. B 95, 024205 (2017).




\bibitem{monthus17mblcayley}
C. Monthus, 
J. Stat. Mech. 123304 (2017).

\bibitem{friedman17}
A. J. Friedman, R. Vasseur, A. C. Potter, S. A. Parameswaran, Phys. Rev. B 98, 064203 (2018)


\bibitem{rsrgxmajorana}
 C. Monthus, 
 J. Phys. A: Math. Theor.  51 115304 (2018).


\bibitem{strong_review}
F. Igl\'oi and C. Monthus, 
Phys. Rep. 412, 277 (2005); \\
F. Igloi and C. Monthus, 
Eur. Phys. J. B  91, 290 (2018).


\bibitem{ma_dasgupta}       
       S.-K. Ma, C. Dasgupta, and C.-k. Hu,
       Phys. Rev. Lett. 43, 1434 (1979) ; \\
      C. Dasgupta and S.-K. Ma
       Phys. Rev. B 22, 1305 (1980).

\bibitem{fisher}
D. S. Fisher,
Phys. Rev. Lett. 69, 534 (1992) ; \\
D. S. Fisher,
Phys. Rev. B 51, 6411 (1995).







\bibitem{emergent_swingle}
B. Swingle, arxiv:1307.0507.

\bibitem{emergent_serbyn}
M. Serbyn, Z. Papic and D.A. Abanin, Phys. Rev. Lett. 111, 127201 (2013).

\bibitem{emergent_huse}
D.A. Huse, R. Nandkishore and V. Oganesyan, Phys. Rev. B 90, 174202 (2014).

\bibitem{emergent_ent}
A. Nanduri, H. Kim and D.A. Huse, Phys. Rev. B 90, 064201 (2014).

\bibitem{imbrie}
J. Z. Imbrie,  J. Stat. Phys. 163, 998 (2016).

\bibitem{serbyn_quench}
M. Serbyn, Z. Papic and D.A. Abanin, Phys. Rev. B 90, 174302 (2014).

\bibitem{emergent_vidal}
A. Chandran, I.H. Kim, G. Vidal and  D.A. Abanin, Phys. Rev. B 91, 085425 (2015).

\bibitem{emergent_ros}
V. Ros, M. M\"uller and A. Scardicchio, Nucl. Phys. B 891, 420 (2015).

\bibitem{emergent_rademaker}
L. Rademaker and M. Ortuno, Phys. Rev. Lett. 116, 010404 (2016).

\bibitem{serbyn_powerlawent}
M. Serbyn, A. A. Michailidis, D. A. Abanin, Z. Papic, Phys. Rev. Lett. 117, 160601 (2016).

\bibitem{c_emergent}
C. Monthus, J. Stat. Mech. 033101 (2016).

\bibitem{ros_remanent}
V. Ros and M. Mueller,  	Phys. Rev. Lett. 118, 237202 (2017)

\bibitem{wortis}
R Wortis and Malcolm P Kennett, J. Phys.: Condens. Matter 29, 405602 (2017)




\bibitem{imbrie17}
J. Z. Imbrie, V. Ros, A. Scardicchio, 
 Annalen der Physik 529, 1600278 (2017).

\bibitem{rademaker17}
L. Rademaker, M. Ortuno and A.M. Somoza, 
 Annalen der Physik 529, 1600322 (2017).


\bibitem{perlioms}
 C. Monthus, 
J. Phys. A: Math. Theor. 51 195301 (2018).

\bibitem{maj_pol}
A. Wieckowski, M. M. Maska, M. Mierzejewski, Phys. Rev. Lett. 120, 040504 (2018)

\bibitem{counting_lioms}
M. Mierzejewski, M. Kozarzewski and P. Prelovsek, 	Phys. Rev. B 97, 064204 (2018)





 \bibitem{vosk13}
R. Vosk and E. Altman, 
Phys. Rev. Lett. 110, 067204 (2013).

 \bibitem{vosk14}
R. Vosk and E. Altman, 
Phys. Rev. Lett. 112, 217204 (2014).

\bibitem{bukov15}
M. Bukov, L. D'Alessio and A. Polkovnikov, 
Advances in Physics, Vol. 64, No. 2, 139 (2015).


 \bibitem{huang17}
Y. Huang, 
Annals of Physics 380, 224 (2017).

\bibitem{monthus17rsrgt}
C. Monthus,
J. Phys. A: Math. Theor. 51 275302 (2018).


\bibitem{mbllandscape}
S. Balasubramanian, Y. Liao and V. Galitski, arxiv:1908.05283.





\bibitem{wolf}
D. Perez-Garcia, F. Verstraete, M. M. Wolf and J. I. Cirac, Quantum Inf. Comput. 7, 401 (2007).


\bibitem{ver}
F. Verstraete, V. Murg, J.I. Cirac,
Advances in Physics 57 (2), 143 (2008).

\bibitem{cirac}
J. I. Cirac, F. Verstraete, J. Phys. A: Math. Theor. 42, 504004 (2009).

\bibitem{vidal_intro}
G. Vidal, " Entanglement Renormalization: an introduction ", 
chapter of the book "Understanding Quantum Phase Transitions", edited by Lincoln D. Carr (Taylor and Francis, Boca Raton, 2010)

\bibitem{ulrich2011}
U. Schollw\"ock, Annals of Physics 326, 96  (2011)

\bibitem{phd-evenbly}
G. Evenbly, PhD thesis "Foundations and Applications of Entanglement Renormalization ", arxiv:1109.5424.

\bibitem{mera-review}
G. Evenbly and G. Vidal, chapter 4 in the book "Strongly Correlated Systems. Numerical Methods", edited by A. Avella and F. Mancini (Springer Series in Solid-State Sciences, Vol. 176 (2013).

\bibitem{hauru}
M. Hauru, Master Thesis "Multiscale Entanglement Renormalisation Ansatz" (2013).

\bibitem{orus14a}
R. Orus, Annals of Physics 349, 117 (2014).

\bibitem{orus14b}
R. Orus, Eur. Phys. J. B  87, 280 (2014).

\bibitem{orus19}
R. Orus, Nature Reviews Physics 1, 538 (2019)


\bibitem{ranard}
X. L. Qi and D. Ranard, Quantum 3, 159 (2019)

\bibitem{clark}
E. Chertkov and B. K. Clark, 	Phys. Rev. X 8, 031029 (2018)

\bibitem{dupont1}
M. Dupont and N. Laflorencie, Phys. Rev. B 99, 020202 (2019)

\bibitem{dupont2}
M. Dupont, N. Mac\'e and N. Laflorencie, Phys. Rev. B 100, 134201 (2019)

\bibitem{topo}
%
E. Chertkov, B. Villalonga, B. K. Clark, arxiv:1910.10165




\bibitem{amico08} 
L. Amico, R. Fazio, A. Osterloh, and V. Vedral, 
Entanglement in many-body systems
Rev. Mod. Phys. 80, 517 (2008).

\bibitem{horo}
R. Horodecki, P. Horodecki, M. Horodecki and K. Horodecki, Rev. Mod. Phys. 81, 865 (2009).

\bibitem{calabrese09} 
P. Calabrese, J. Cardy and B. Doyon (Eds. of the special issue)
Entanglement entropy in extended quantum systems,
 J. Phys. A 42 500301 (2009).


\bibitem{qi}
B. Zeng, X. Chen, D.L. Zhou, X.G. Wen
"Quantum Information Meets Quantum Matter : From Quantum Entanglement to Topological Phase in Many-Body Systems", arxiv:1508.02595.


\bibitem{laflorencie16} 
 N. Laflorencie, Phys. Rep. 643, 1 (2016).

\bibitem{chiara}
G. De Chiara and A. Sanpera, Report on Prog. Phys. 81 074002 (2018).


\bibitem{hastings}
M. B. Hastings, JSTAT, P08024 (2007).

\bibitem{vidalcanonical} 
G. Vidal, Phys. Rev. Lett. 91, 147902 (2003)




\bibitem{jms2009}
J.M. Stephan, S. Furukawa, G. Misguich and V. Pasquier, Phys. Rev. B 80, 184421 (2009).

\bibitem{jms2010}
J.M. Stephan,  G. Misguich and V. Pasquier, Phys. Rev. B 82, 125455 (2010).

\bibitem{jms2011}
J.M. Stephan,  G. Misguich and V. Pasquier, Phys. Rev. B 84, 195128 (2011).

\bibitem{moore}
M.P. Zaletel, J.H. Bardarson and J.E. Moore, Phys. Rev. Lett. 107, 020402 (2011).

\bibitem{atas_short}
Y.Y. Atas and E. Bogomolny, Phys. Rev. E 86, 021104 (2012).

\bibitem{grassberger}
H.W. Lau and P. Grassberger, Phys. Rev. E 87,022128 (2013).

\bibitem{atas_long}
Y.Y. Atas and E. Bogomolny, Phil. Trans. R. Soc. A 372, 20120520 (2014).

\bibitem{luitz_short}
D.J. Luitz, F. Alet and N. Laflorencie, Phys. Rev. Lett. 112, 057203 (2014).

\bibitem{luitz_o3}
D.J. Luitz, F. Alet and N. Laflorencie, Phys. Rev. B 89, 165106 (2014).

\bibitem{luitz_spectro}
D.J. Luitz, F. Alet and N. Laflorencie, J. Stat. Mech. P08007 (2014).

\bibitem{luitz_qmc}
D.J. Luitz, X. Plat, N. Laflorencie and F. Alet, Phys. Rev. B 90, 125105 (2014).

\bibitem{jms2014}
J.M. Stephan, Phys. Rev. B 90, 045424 (2014).

\bibitem{alcaraz1}
F.C. Alcaraz and M.A. Rajabpour, Phys. Rev. B 90, 075132 (2014).

\bibitem{alcaraz2}
F.C. Alcaraz and M.A. Rajabpour, Phys. Rev. B 91, 155122 (2015).

\bibitem{c_renyi}
C. Monthus, J. Stat. Mech. P04007 (2015).

\bibitem{jms2017}
M. Brockmann and J.M. Stephan, J. Phys. A: Math. Theor. 50, 354001 (2017).

\bibitem{c_treetensor}
C. Monthus, J. Phys. A: Math. Theor. 51 095301 (2018).




\bibitem{mace1}
N. Mac\'e, N. Laflorencie and F. Alet, SciPost Phys. 6, 050 (2019)

\bibitem{mace2}
N. Mac\'e, F. Alet and N. Laflorencie, arXiv:1812.10283

\bibitem{lev}
D. J. Luitz, I. M. Khaymovich and Y. Bar Lev, arXiv:1909.06380



\bibitem{kitaevchain}
A.Y. Kitaev, Phys. Usp. 44, 131 (2011).

\bibitem{kitaevfid}
L. Fidkowski and A. Kitaev, Phys. Rev. B 83, 075103 (2011).


\bibitem{kitaevalpha}
R. Verresen, R. Moessner and F. Pollmann, Phys. Rev. B 96, 165124 (2017).

\bibitem{10phases}
B. Friedman, A. Rajak, A. Russomanno and E.G. Dalla Torre, Europhysics Letters 125, 10008 (2019).



\bibitem{strongzeromode}
P. Fendley, J. Phys. A: Math. Theor. 49 (2016) 30LT01

\bibitem{longcoherence}
J. Kemp, N.Y. Yao, C.R. Laumann and P. Fendley,  J. Stat. Mech.  063105 (2017).

\bibitem{mila}
I.A. Maceira and F. Mila, 	Phys. Rev. B 97, 064424 (2018)

\bibitem{prethermal}
D. Else, P. Fendley, J. Kemp and C. Nayak, 	Phys. Rev. X 7, 041062 (2017)

\bibitem{carollo1}
L.M. Vasiloiu, F. Carollo, J. P. Garrahan, Phys. Rev. B 98, 094308 (2018)

\bibitem{carollo2}
L. M. Vasiloiu, F. Carollo, M. Marcuzzi, J. P. Garrahan
Phys. Rev. B 100, 024309 (2019)

\bibitem{akhmerov}
A.R. Akhmerov, Phys. Rev. B 82, 020509(R) (2010).

\bibitem{goldstein}
G. Goldstein and C. Chamon, Phys. Rev. B 86, 115122 (2012).

\bibitem{wilczek}
J. Lee and F. Wilczek, Phys. Rev. Lett. 111, 226402(2013).

\bibitem{feldman}
G. Yang, D. E. Feldman, 	Phys. Rev. B 89, 035136 (2014).

\bibitem{moreabout}
Li-Wei Yu, Mo-Lin Ge, Scientific Reports 5, 8102 (2015).

\bibitem{kauffman}
R. Ul Haq, L. H. Kauffman, arxiv:1704.00252.

\bibitem{c_evenodd}
C. Monthus, J. Phys. A: Math. Theor. 51 265303 (2018).



\bibitem{mpdo}
F. Verstraete, J. J. Garcia-Ripoll, and J. I. Cirac
Phys. Rev. Lett. 93, 207204 (2004)

\bibitem{mpd}
M. Zwolak and G. Vidal, Phys. Rev. Lett. 93, 207205 (2004).



\bibitem{prosen}
T. Prosen and I. Pizorn, Phys. Rev. A 76, 032316 (2007)

\bibitem{zhou}
T. Zhou and D. J. Luitz,
Phys. Rev. B 95, 094206 (2017).

\bibitem{whiteR}
C. D. White, M. Zaletel, R. S. K. Mong, G. Refael, Phys. Rev. B 97, 035127 (2018).

\bibitem{zilber}
E. van Nieuwenburg and O. Zilberberg, Phys. Rev. A 98, 012327 (2018)



\end{thebibliography}
\end{document}